\documentclass[aps,pra,twocolumn,superscriptaddress,showpacs]{revtex4-1}
\usepackage{amsfonts,amssymb,amsmath,mathrsfs}
\usepackage{graphics,graphicx,epsfig}
\usepackage{color}
\usepackage{bm,bbm}
\usepackage{braket}
\usepackage{rotating}
\newcommand{\cB}{\mathcal{B}}
\newcommand{\cD}{\mathcal{D}}
\newcommand{\cE}{\mathcal{E}}
\newcommand{\cH}{\mathcal{H}}
\newcommand{\cR}{\mathcal{R}}
\newcommand{\cT}{\mathcal{T}}
\newcommand{\cV}{\mathcal{V}}
\newcommand{\id}{\mathbbm{1}}
\newcommand{\tr}{\textrm{tr}}
\newcommand{\upe}{\textrm{e}}
\newcommand{\upi}{\textrm{i}}
\newcommand{\upd}{\textrm{d}}
\newcommand{\vc}{\textrm{vec}}
\newcommand{\rrangle}{\rangle\!\rangle}
\newcommand{\llangle}{\langle\!\langle}
\newcommand{\STP}{\mathscr{S}_{\mathrm{TP}}}
\newcommand{\Favg}{F_{\mathrm{avg}}}
\newcommand{\Fmin}{F_{\mathrm{min}}}
\newcommand{\RBR}{\textrm{RBR}}

\usepackage[normalem]{ulem}
\newcommand{\SetGrey}[1]{\definecolor{Grey}{gray}{#1}}
\SetGrey{0.60}
\newcommand{\Expi}[1]{\upe^{\upi{#1}}}
\newcommand{\sfrac}[2]%
   {\frac{\mbox{\footnotesize$#1$}}{\mbox{\footnotesize$#2$}}}
\newcommand{\KeyWords}[1]{\newline\rule{0em}{16pt}%
  {\footnotesize{Keywords:\hfill\begin{minipage}[t]{358pt}#1\end{minipage}}}}
\newcommand{\FigOne}{%
\begin{figure}
\centerline{\includegraphics[viewport=62 430 292 738]{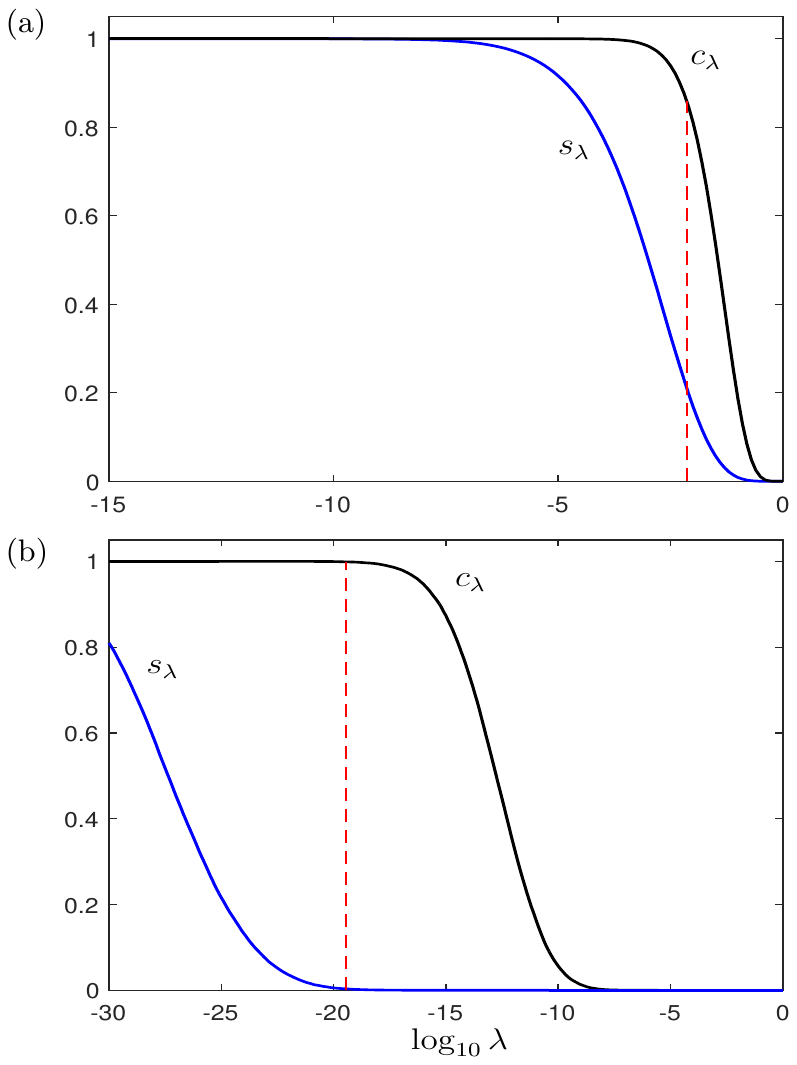}}
\caption{\label{fig:SCR}%
  Size $s_\lambda$ and credibility $c_\lambda$ of the BLRs $\cR_\lambda$,
  plotted against $\log_{10}\lambda$, for (a) the qubit amplitude-damping
  channel, and (b) the qutrit amplitude-damping channel.
  The red vertical dashed lines mark the respective critical $\lambda$ values, 
  ${\lambda_{\text{crit}}=0.0073}$ for (a), and
  ${\lambda_{\text{crit}}=3.5598\times10^{-20}}$ for (b).
  These identify the plausible regions. } 
\end{figure}%
}

\newcommand{\FigTwo}{%
\begin{figure}
\centerline{\includegraphics[viewport=62 555 288 738]{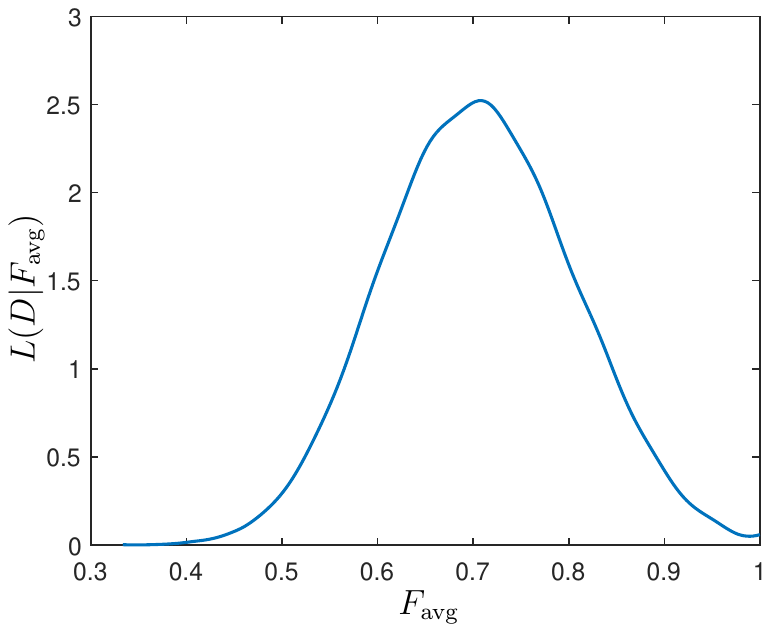}}
\caption{\label{fig:avgMarginalLikelihood}%
The marginal likelihood $L(D|\Favg)$, computed using the iterative
procedure of Ref.~\cite{OEI16} and HMC with our channel
parameterization.}
\end{figure}%
}

\newcommand{\FigThree}{%
\begin{figure}
\centerline{\includegraphics[viewport=62 432 292 725]{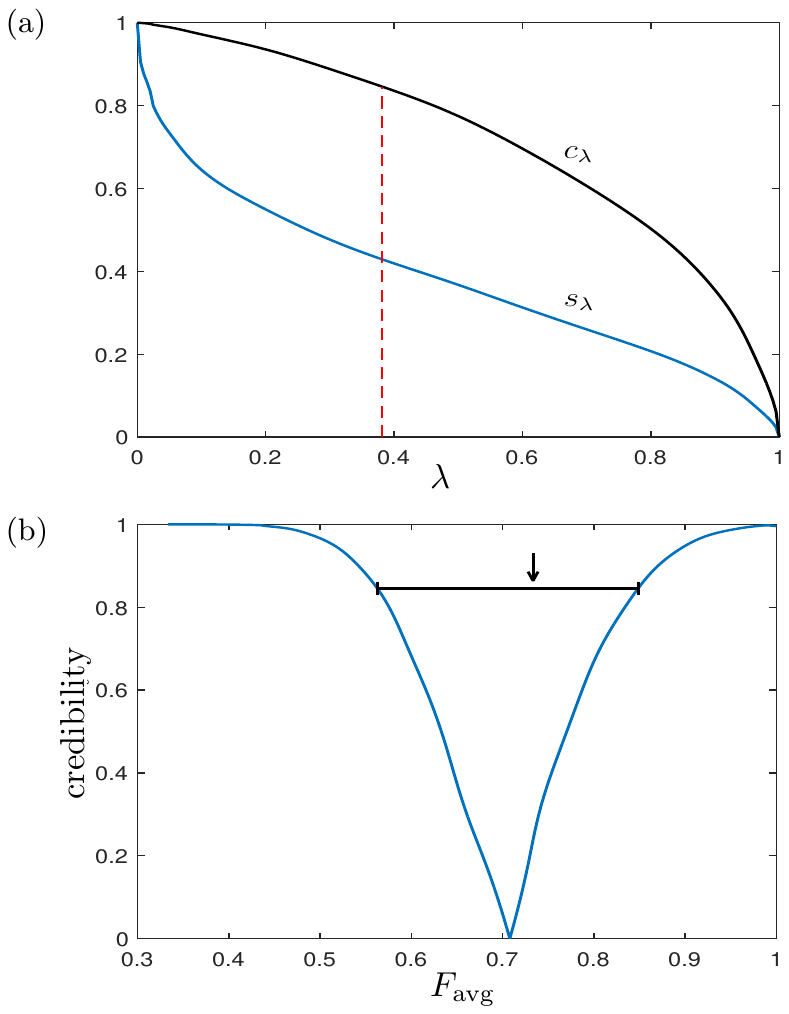}}
\caption{\label{fig:OEI}%
(a) Size (blue) and credibility (green) curves for the bounded likelihood
intervals for $\Favg$. The red vertical dashed line marks the critical value
of $\lambda$, at ${\lambda_{\text{crit}}=0.3819}$.
(b) SCI for $\Favg$. The blue curve indicates the boundaries of the
SCIs for different credibility values. The black horizontal line marks the
plausible interval and the arrow indicates the true value of
${\Favg=0.7333}$.}  
\end{figure}%
}

\newcommand{\FigFour}{%
\begin{figure}
\centerline{\includegraphics[viewport=62 555 288 738]{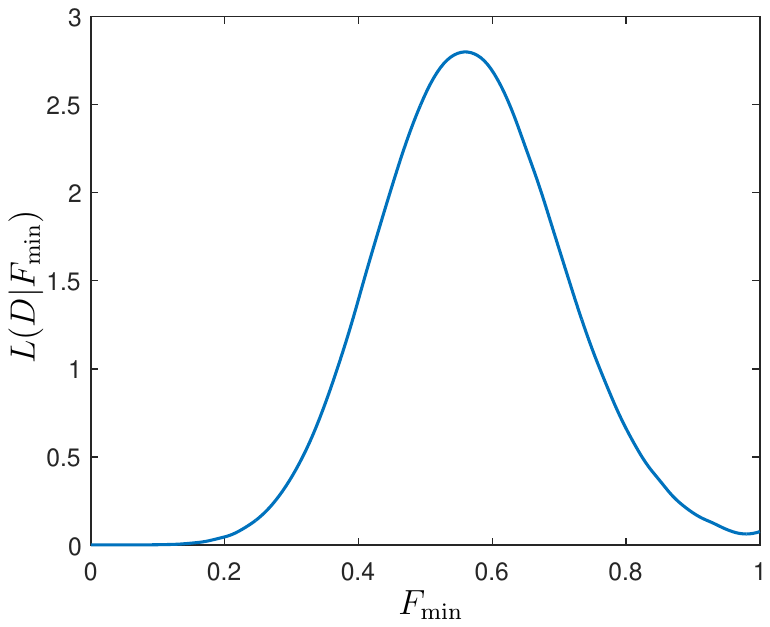}}
\caption{\label{fig:minMarginalLikelihood}
The marginal likelihood $L(D|\Fmin)$, computed using the iterative
procedure of Ref.~\cite{OEI16} and HMC with our channel
parameterization.} 
\end{figure}%
}

\newcommand{\FigSix}{%
\begin{figure}
\centerline{\includegraphics[viewport=62 210 288 740]{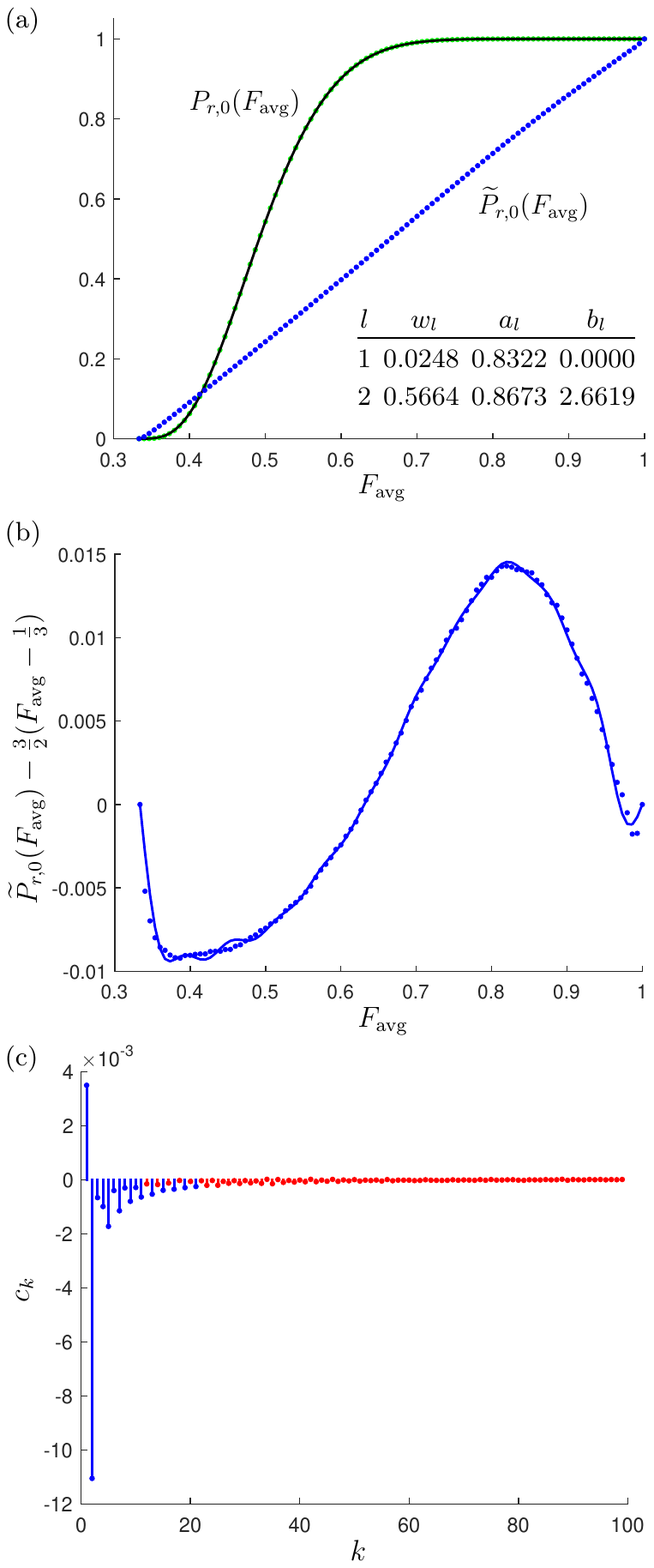}}
\caption{\label{fig:avgGateFid}%
  Average gate fidelity.
  (a) The green dots depict the MC values of $P_{r,0}(\Favg)$, and the
  black curve is fitted to them; the inset table reports the fitting
  parameters.
  The values of $\tilde P_{r,0}(\Favg)$ are traced out by the blue dots.
  (b) The blue dots show the MC-values of $\tilde P_{r,0}(\Favg)$
  after subtracting the straight line ${\frac{3}{2}(\Favg-\frac{1}{3})}$.
  The blue curve, a truncated Fourier series, is fitted to the dots.
  (c) Fourier amplitudes for $\tilde P_{r,0}(\Favg)$.
  The high-frequency noise is removed from the fit in (b) by discarding the
  red amplitudes whose magnitude is less than $2\%$ of 
  that of largest amplitude.}
\end{figure}%
}

\newcommand{\FigSeven}{%
\begin{figure}
\centerline{\includegraphics[viewport=62 210 288 740]{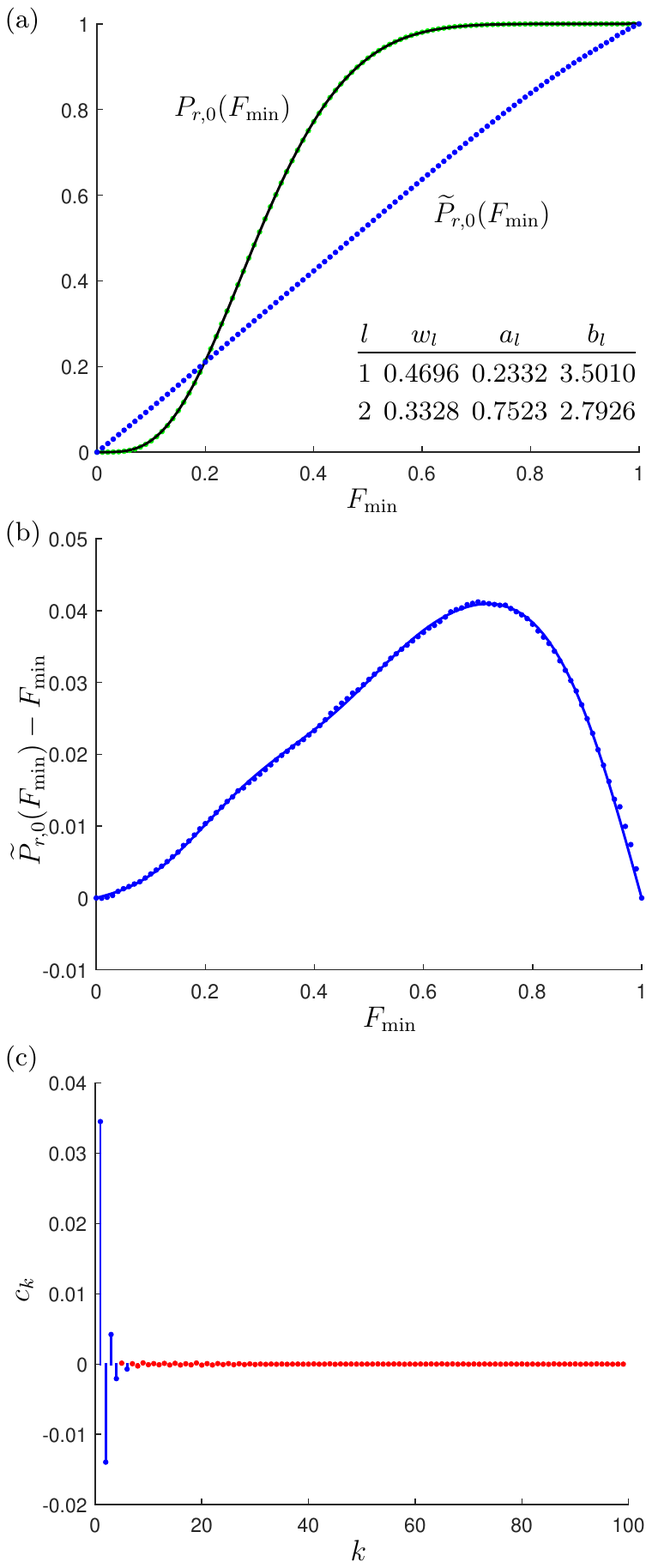}}
\caption{\label{fig:minGateFid}%
  Minimum gate fidelity.
  (a) The green dots depict the MC values of $P_{r,0}(\Fmin)$, and the
  black curve is fitted to them; the inset table reports the fitting
  parameters. 
  The values of $\tilde P_{r,0}(\Fmin)$ are traced out by the blue dots.
  (b) The blue dots show the MC-values of $\tilde P_{r,0}(\Fmin)$
  after subtracting the straight line $\Fmin$.
  The blue curve, a truncated Fourier series, is fitted to the dots.
  (c) Fourier amplitudes for $\tilde P_{r,0}(\Fmin)$.
  The high-frequency noise is removed from the fit in (b) by discarding the
  red amplitudes whose magnitude is less than $2\%$ of 
  that of largest amplitude.} 
\end{figure}%
}

\newcommand{\HowMany}[1]{\begin{turn}{90}%
  \makebox[0pt][l]{\,\underline{\makebox[195pt][c]{$N={#1}$}}}\end{turn}
  &\begin{turn}{90}\makebox[0pt][l]{%
      \rule{2pt}{0pt}\underline{\rule{16pt}{0pt}RBR\rule{16pt}{0pt}}
      \rule{14pt}{0pt}  
   \underline{\rule{18pt}{0pt}BIC\rule{18pt}{0pt}}\rule{16pt}{0pt}%
   \underline{\rule{18pt}{0pt}AIC\rule{18pt}{0pt}}}\end{turn}}

\newcommand{\TblOne}{%
\begin{table*}
  \caption{\label{tbl:ModelSelectionResult}%
   Comparison of results based on AIC, BIC, and RBR criteria, with different
   number of measured copies $N$.
   Candidate models are nested channel families (see main text).
   Each row below collects the counts for each family of true channels; each
   column collects the counts for the model that a criterion selects as the
   best fit for the data.} 
 \hspace*{-1.7cm}{\scalebox{0.968}{%
\begin{minipage}[c]{0.4\textwidth}
   \vspace*{0.1cm}
  \begin{tabular}{ll@{~}c@{\!}c@{\!\!}c@{~~}c@{~~}c@{~~}c@{~~}c}\hline\hline
    &&true&& \multicolumn{5}{c}{\# cases where the %
                                best-fit model is} \\[-0.5ex]
  &&family&\quad\quad& Dephasing & Pauli & SUnital & Unital & General  \\
\hline
  &&Dephasing   && 947 & 43 & 10 & 0 & 0 \\
  &&Pauli      && 583 & 408 & 9 & 0 & 0 \\ 
  &&SUnital  && 629 & 319 & 52 & 0 & 0 \\
  &&Unital  && 562 & 405 & 31 & 2 & 0 \\
  &&General  && 596 & 372 & 30 & 2 & 0 \\[-2ex]  
  &&&&\multicolumn{5}{c}{\hrulefill}\\
  &&Dephasing   && 983 & 17 & 0 & 0 & 0 \\
  &&Pauli      && 721 & 279 & 0 & 0 & 0 \\
  &&SUnital  && 795 & 200 & 5 & 0 & 0 \\
  &&Unital  && 741 & 256 & 3 & 0 & 0 \\
  &&General  && 762 & 235 & 3 & 0 & 0 \\[-2ex]
  &&&&\multicolumn{5}{c}{\hrulefill}\\
  &&Dephasing   && 712 & 62 & 93 & 56 & 77 \\
  &&Pauli      && 224 & 271 & 173 & 140 & 192 \\ 
  &&SUnital  && 239 & 151 & 315 & 117 & 178 \\
  &&Unital  && 191 & 164 & 215 & 234 & 196 \\
   \HowMany{20}%
  &General  && 166 & 156 & 180 & 164 & 334 \\[1ex] 
  \hline
  &&Dephasing   && 935 & 42 & 21 &  2 & 0 \\
  &&Pauli      && 305 & 644 & 45 & 6 & 0 \\
  &&SUnital  && 343 & 429 & 213 & 13 & 2 \\
  &&Unital  && 257 & 534 & 148 & 57 & 4 \\
  &&General  && 278 & 531 & 115 & 33 & 43 \\[-2ex]  
  &&&&\multicolumn{5}{c}{\hrulefill}\\
  &&Dephasing   && 995 & 4 & 1 & 0 & 0 \\
  &&Pauli      && 539 & 461 &  0 & 0 & 0 \\ 
  &&SUnital  && 646 & 328 & 26 & 0 & 0 \\
  &&Unital  && 601 & 393 & 6 & 0 & 0 \\
  &&General  && 586 & 404 & 10 & 0 & 0 \\[-2ex]
  &&&&\multicolumn{5}{c}{\hrulefill}\\
  &&Dephasing   && 823 & 52 & 68 & 22 & 35 \\
  &&Pauli      && 158 & 404 & 139 & 129 & 170 \\
  &&SUnital  && 155 & 197 & 376 & 140 & 132 \\
  &&Unital  && 91 & 185 & 222 & 343 & 159 \\
   \HowMany{50}%
  &General  && 75 & 178 & 139 & 176 & 432 \\[1ex]  
  \hline
  &&Dephasing   && 938 & 41 & 18 & 1 & 2 \\
  &&Pauli      && 173 & 733 & 72 & 16 & 6 \\
  &&SUnital  && 141 & 368 & 455 & 30 & 6 \\
  &&Unital  && 87 & 409 & 264 & 215 & 25 \\
  &&General  && 78 & 442 & 147 & 93 & 240 \\[-2ex]  
  &&&&\multicolumn{5}{c}{\hrulefill}\\
  &&Dephasing   && 998 & 2 & 0 & 0 & 0 \\
  &&Pauli      && 367 & 631 & 2 & 0 & 0 \\ 
  &&SUnital  && 427 & 471 & 102 & 0 & 0 \\
  &&Unital  && 357 & 593 & 42 & 8 & 0 \\
  &&General  && 363 & 609 & 26 & 2 & 0 \\[-2ex]
  &&&&\multicolumn{5}{c}{\hrulefill}\\
  &&Dephasing   && 905 & 43 & 33 & 11 & 8 \\
  &&Pauli      && 129 & 578 & 105 & 94 & 94 \\
  &&SUnital  && 76 & 210 & 506 & 126 & 82 \\
  &&Unital  && 55 & 196 & 223 & 394 & 132 \\
     \HowMany{100}%
  &General  && 34 & 132 & 118 & 190 & 526 \\[1ex]
  \hline\hline
  \end{tabular}
  \end{minipage}\hspace*{2cm}%
  \begin{minipage}[c]{0.4\textwidth}
   \vspace*{0.1cm}
  \begin{tabular}{ll@{~}c@{\!}c@{\!\!}c@{~~}c@{~~}c@{~~}c@{~~}c}\hline\hline
    &&true&& \multicolumn{5}{c}{\# cases where the %
                                   best-fit model is}\\[-0.5ex]
  &&family&\quad\quad& Dephasing & Pauli & SUnital & Unital & General  \\
  \hline
  &&Dephasing   && 933 & 40 & 18 & 7 & 2 \\ 
  &&Pauli      && 22 & 866 & 79 & 26 & 7 \\
  &&SUnital  && 0 & 64 & 818 & 87 & 31 \\
  &&Unital  && 0 & 10 & 85 & 811 & 94 \\
  &&General  && 0 & 2 & 3 & 47 & 948 \\[-2ex]  
  &&&&\multicolumn{5}{c}{\hrulefill}\\
  &&Dephasing   && 1000 & 0 & 0 & 0 & 0 \\ 
  &&Pauli      && 77 & 923 & 0 & 0 & 0 \\
  &&SUnital  && 27 & 231 & 742 & 0 & 0 \\
  &&Unital  && 3 & 179 & 250 & 568 & 0 \\
  &&General  && 2 & 113 & 105 & 112 & 668 \\[-2ex] 
    &&&&\multicolumn{5}{c}{\hrulefill}\\
    &&Dephasing   && 987 & 12 & 1 & 0 & 0 \\ 
  &&Pauli      && 36 & 936 & 20 & 8 & 0 \\
  &&SUnital  && 1 & 92 & 864 & 35 & 8 \\ 
  &&Unital  && 0 & 18 & 122 & 837 & 23 \\
   \HowMany{1\,000}%
  &General  && 0 & 4 & 7 & 106 & 883 \\[1ex] 
  \hline
  &&Dephasing   && 911 & 49 & 29 & 6 & 5 \\
  &&Pauli      && 2 & 846 & 99 & 34 & 19 \\
  &&SUnital  && 0 & 1 & 868 & 94 & 37 \\
  &&Unital  && 0 & 0 & 3 & 889 & 108 \\
  &&General  && 0 & 0 & 0 & 1 & 999 \\[-2ex]  
  &&&&\multicolumn{5}{c}{\hrulefill}\\
  &&Dephasing   && 1000 & 0 & 0 & 0 & 0 \\
  &&Pauli      && 11 & 989 & 0 & 0 & 0 \\
  &&SUnital  && 0 & 11 & 989 & 0 & 0 \\
  &&Unital  && 0 & 0 & 37 & 963 & 0 \\
  &&General  && 0 & 0 & 0 & 17 & 983 \\[-2ex]
    &&&&\multicolumn{5}{c}{\hrulefill}\\
    &&Dephasing   && 999 & 1 & 0 & 0 & 0 \\
  &&Pauli      && 6 & 993 & 1 & 0 & 0 \\
  &&SUnital  && 0 & 7 & 985 & 8 & 0 \\
  &&Unital  && 0 & 2 & 60 & 919 & 19 \\
   \HowMany{10\,000}%
  &General  && 0 & 0 & 7 & 86 & 907 \\[1ex]  
  \hline
  &&Dephasing   && 921 & 44 & 27 & 6 & 2 \\
  &&Pauli      && 1 & 848 & 97 & 37 & 17 \\
  &&SUnital  && 0 & 0 & 865 & 102 & 33 \\
  &&Unital  && 0 & 0 & 0 & 898 & 102 \\
  &&General  && 0 & 0 & 0 & 0 & 1000 \\[-2ex]  
  &&&&\multicolumn{5}{c}{\hrulefill}\\
  &&Dephasing   && 1000 & 0 & 0 & 0 & 0 \\
  &&Pauli      && 2 & 998 & 0 & 0 & 0 \\
  &&SUnital  && 0 & 1 & 999 & 0 & 0 \\
  &&Unital  && 0 & 0 & 1 & 999 & 0 \\
  &&General  && 0 & 0 & 0 & 1 & 999 \\[-2ex]
    &&&&\multicolumn{5}{c}{\hrulefill}\\
    &&Dephasing   && 1000 & 0 & 0 & 0 & 0 \\
  &&Pauli      && 1 & 999 & 0 & 0 & 0 \\
  &&SUnital  && 0 & 4 & 994 & 2 & 0 \\
  &&Unital  && 0 & 0 & 68 & 924 & 8 \\
     \HowMany{100\,000}%
  &General  && 0 & 0 & 7 & 80 & 913 \\[1ex] 
  \hline\hline
  \end{tabular}
  \end{minipage}}}%
\end{table*}}

\newcommand{\ThisMany}[1]{\begin{turn}{90}%
    \makebox[0pt][l]{\underline{\makebox[58pt][c]{$N={#1}$}}}%
  \end{turn}}

\newcommand{\TblTwo}{%
\begin{table}
  \caption{\label{tbl:Bias}%
    A check for bias in the prior.
    1000 random channels from each of the channel families are drawn, and data
    for  with different number of measured copies $N$ are simulated for each
    true channel.
    The table shows the fraction of instances with evidence against each of
    the channel families.} 
\centerline{%
  \begin{tabular}{l@{~~}c@{\!}c@{\!\!}c@{~~}c@{~~}c@{~~}c@{~~}c}\hline\hline
  &true&& \multicolumn{5}{c}{fraction with evidence against}\\[-0.5ex]
  &family&\quad\quad& Dephasing & Pauli & SUnital & Unital & General  \\
\hline
  &Dephasing   && 0.233 & 0.812 & 0.746 & 0.798 & 0.810 \\
  &Pauli      &&  0.724 & 0.413 & 0.531 & 0.401 & 0.495 \\
  &SUnital  &&   0.687 & 0.558 & 0.398 & 0.455 & 0.515 \\
  &Unital  &&   0.767 & 0.518 & 0.508 & 0.316 & 0.406 \\
    \ThisMany{20}
    &General  &&  0.779 & 0.524 & 0.552 & 0.384 & 0.350 \\[-1ex] 
  &&&\multicolumn{5}{c}{\hrulefill}\\
  &Dephasing   && 0.127 & 0.802 & 0.827 & 0.911 & 0.916 \\
  &Pauli      &&  0.786 & 0.322 & 0.588 & 0.511 & 0.638 \\
  &SUnital  && 0.785 & 0.578 & 0.326 & 0.454 & 0.642 \\ 
  &Unital  && 0.876 & 0.561 & 0.503 & 0.276 & 0.488 \\ 
  \ThisMany{50}&General  && 0.892 & 0.610 & 0.658 & 0.448 & 0.311 \\[-1ex] 
  &&&\multicolumn{5}{c}{\hrulefill}\\
  &Dephasing   && 0.042 & 0.844 & 0.911 & 0.964 & 0.978 \\
  &Pauli      &&  0.826 & 0.228 & 0.623 & 0.672 & 0.780 \\
  &SUnital  && 0.874 & 0.613 & 0.249 & 0.508 & 0.793 \\ 
  &Unital  && 0.925 & 0.658 & 0.573 & 0.260 & 0.596 \\ 
  \ThisMany{100}&General  && 0.948 & 0.715 & 0.731 & 0.576 & 0.266 \\[-1ex] 
  &&&\multicolumn{5}{c}{\hrulefill}\\
  &Dephasing   && 0.003 & 0.966 & 0.998 & 1 & 1 \\ 
  &Pauli      && 0.946 & 0.028 & 0.930 & 0.981 & 0.999 \\
  &SUnital  && 0.992 & 0.878 & 0.088 & 0.868 & 0.983 \\
  &Unital  && 1 & 0.974 & 0.821 & 0.088 & 0.913 \\
  \ThisMany{1\,000}&General  && 1 & 0.995 & 0.983 & 0.868 & 0.077 \\[-1ex] 
  &&&\multicolumn{5}{c}{\hrulefill}\\
  &Dephasing   && 0.001 & 0.996 & 1 & 1 & 1 \\
  &Pauli      && 0.992 & 0.006 & 0.996 & 1 & 1 \\
  &SUnital  && 1 & 0.992 & 0.011 & 0.989 & 1 \\
  &Unital  && 1 & 0.998 & 0.930 & 0.073 & 0.981 \\
  \ThisMany{10\,000}&General  && 1 & 1 & 0.991 & 0.907 & 0.092 \\[-1ex] 
  &&&\multicolumn{5}{c}{\hrulefill}\\
  &Dephasing   && 0 & 0.999 & 1 & 1 & 1 \\
  &Pauli      && 0.999 & 0.001 & 1 & 1 & 1 \\ 
  &SUnital  && 1 & 0.996 & 0.006 & 0.998 & 1 \\
  &Unital  && 1 & 1 & 0.932 & 0.076 & 0.992 \\
  \ThisMany{100\,000}&General  && 1 & 1 & 0.993 & 0.920 & 0.086 \\ [1ex]
\hline\hline
  \end{tabular}
}
\end{table}}

\begin{document}
\title{User-specified random sampling of quantum channels and its applications}
\author{Jun Yan Sim}
\email{e0012429@u.nus.edu}
\affiliation{Centre for Quantum Technologies, %
             National University of Singapore, %
             3 Science Drive 2, Singapore 117543, Singapore}

\author{Jun Suzuki}\email{junsuzuki@uec.ac.jp}
\affiliation{Graduate School of Informatics and Engineering, %
  The University of Electro-Communications, 1-5-1 Chofugaoka, Chofu-shi, %
  Tokyo, 182-8585 Japan}

\author{Berthold-Georg Englert}\email{cqtebg@nus.edu.sg}
\affiliation{Centre for Quantum Technologies, %
             National University of Singapore, %
             3 Science Drive 2, Singapore 117543, Singapore}
\affiliation{Department of Physics, National University of Singapore, %
             2 Science Drive 3, Singapore 117551, Singapore}
\affiliation{\mbox{MajuLab, International Joint Research Unit UMI 3654, %
    CNRS, Universit{\'e} C{\^o}te d'Azur, Sorbonne Universit{\'e}}, %
  National University of Singapore, Nanyang Technological University, Singapore}

\author{Hui Khoon Ng}
\email{huikhoon.ng@yale-nus.edu.sg}
\affiliation{Yale-NUS College, 16 College Avenue West, %
             Singapore 138527, Singapore}
\affiliation{Centre for Quantum Technologies, %
             National University of Singapore, %
             3 Science Drive 2, Singapore 117543, Singapore}
\affiliation{\mbox{MajuLab, International Joint Research Unit UMI 3654, %
    CNRS, Universit{\'e} C{\^o}te d'Azur, Sorbonne Universit{\'e}}, %
  National University of Singapore, Nanyang Technological University, Singapore}

\begin{abstract}
Random samples of quantum channels have many applications in quantum
information processing tasks. Due to the Choi--Jamio\l{}kowski isomorphism,
there is a well-known correspondence between channels and states, and one can
imagine adapting \emph{state} sampling methods to sample quantum
channels.
Here, we discuss such an adaptation, using the Hamiltonian Monte
Carlo method, a well-known classical method capable of producing
high-quality samples from arbitrary, user-specified distributions.
Its implementation requires an exact parameterization of the space of quantum
channels, with no superfluous parameters and no constraints.
We construct such a parameterization, and demonstrate its use in three common
channel sampling applications.
\end{abstract}


\maketitle

\section{Introduction}

Quantum channels, or completely positive (CP) and trace-preserving (TP) maps,
are a central concept in describing the dynamics of quantum systems. They form
the basic models for imperfect quantum operations used for quantum information
processing (QIP). Random---according to some specified distribution---samples
of quantum channels are needed in many QIP tasks, including the evaluation of
the distributional average of channel-related quantities, the computation of
error bars for quantum process tomography, the exploration of typical
properties of quantum channels, the numerical optimization of functions of
channels over a complicated landscape, and others.

Sampling from specific distributions over the quantum \emph{state} space is a
well-studied problem, with many different approaches, including the Monte
Carlo (MC) technique for arbitrary distributions, and other methods for
sampling from specific distributions
\cite{Shang+1:15,Seah+1:15,Zyczkowski98,Zyczkowski99,Zyczkowski01,%
  Blume10,Huszar12,YS19}. 
Due to the Choi--Jamio\l{}kowski isomorphism \cite{Choi,Jamiolkowski}, which
gives a correspondence between CP channels and states, these state sampling
methods can be adapted to sample quantum channels.
Indeed, in the recent work by Thinh
\textit{et al.}~\cite{Thinh18}, a Metropolis--Hasting
Markov chain (MHMC) MC approach was used to sample
channels from arbitrary distributions, by sampling the purification of the
Choi--Jamio\l{}kowski state corresponding to the channel.
References \cite{Geometry} and \cite{Bruzda09} discuss a procedure for generating samples of quantum channels with a specific distribution on the channel space by making use of the channel-state correspondence. An alternative way of generating the same distribution of channels is to couple the input state and an ancilla initially in an arbitrary pure state by a Haar-random unitary operator and then taking the partial trace over the ancilla. See Ref. \cite{Bruzda10} for more discussion on the properties of the distribution of channels generated by these two procedures.

As a general method for sampling from arbitrary distributions,
MC methods stand out in their wide-ranging applicability and efficiency.
The MHMC variety of MC methods used, for example, in Ref.~\cite{Thinh18},
however, suffer from strong correlations between sample points, and one
requires large samples for reliable answers not biased by these correlations.
This was observed, for instance, in the MHMC state sampling algorithm
of Ref.~\cite{Shang+1:15}.
A significant improvement in the quality of the samples was seen when we
switched to the Hamiltonian Monte Carlo (HMC) approach \cite{Seah+1:15},
reaffirming the advantage of HMC over MHMC\thinspace MC also observed
in other settings \cite{Neal96, Hajian07, Porter14, Neal11, Duane87}. 

The HMC method requires the availability of a parameterization of the domain
space with exactly the right number of parameters, with no superfluous
parameters and no constraints. 
The parameterization of the channel/state space used in
Ref.~\cite{Thinh18}, which has superfluous parameters, cannot be used
for HMC.
The exact parameterization of states used in the HMC algorithm in
Ref.~\cite{Seah+1:15} gives, through the Choi--Jamio\l{}kowski isomorphism, a 
parameterization of the set of all CP, but not necessarily TP, maps.
The TP property has to be imposed as an explicit constraint, thus rendering
the parameterization unsuitable in a HMC algorithm for sampling CPTP
channels. 

In this work, we construct one exact parameterization of the space of CPTP
maps, with no superfluous parameters, and no constraints.
This can then be used in a HMC procedure for sampling from arbitrary,
user-specified, distributions over the channel space.
To illustrate the usefulness of our parameterization and the HMC algorithm, we  
apply our methods to three quantum sampling problems.
Our examples are focused on problems in quantum process tomography, reflecting
the interests of the authors; our parameterization and the HMC method,
however, are just as useful for sampling problems in other areas of
QIP.
As an aside, our construction exactly parameterizes the space of all
bipartite mixed quantum states with the completely mixed state for one of the
parties.

Here is the brief outline of our paper.
We first review the Choi--Jamio\l{}kowski isomorphism in
Sec.~\ref{sec:isomorphism}.
Section~\ref{sec:param} explains our main contribution: the
exact parameterization of the space of CPTP channels.
In Sec.~\ref{sec:applications}, we illustrate the use of our parameterization
in a HMC sampling algorithm through three examples from quantum process
tomography: (a) the construction of error regions in process estimation; (b)
marginal likelihood for estimating specific properties of the channel; (c)
model selection among candidate channel families.
The reader is referred to Ref.~\cite{Seah+1:15} or Appendix~\ref{app:HMC} 
for an introduction to the HMC algorithm used here.
We conclude in Sec.~\ref{sec:conc}.

\section{The channel-state duality}\label{sec:isomorphism}
There are many ways of writing the CPTP map of a quantum channel.
Given our desire to make the connection with the sampling of quantum
states, we make use of the channel-state duality and describe the quantum
channel by a state via the Choi--Jamio\l{}kowski isomorphism.
Here, we remind the reader of this isomorphism, and, in the process, define
the notation used throughout the article.

We begin with the $d$-dimensional Hilbert space $\cH$ describing the state
vectors (pure states) of the system. We define a map $*:\cH\rightarrow\cH$, 
\begin{equation}
  *(|\psi\rangle)\equiv|\overline\psi\rangle\in\cH,
  \quad \textrm{for }|\psi\rangle\in\cH,
\end{equation}
such that 
\begin{equation}\label{eq:sym}
  \langle\overline\psi|\overline\phi\rangle
  =\langle\phi|\psi\rangle,\quad\forall|\psi\rangle,|\phi\rangle\in\cH,
\end{equation}
and $*$ is ``$*$-linear'', i.e.,
\begin{equation}
  *\Bigl(\sum_ic_i|\psi_i\rangle\Bigr)=\sum_ic_i^*|\overline\psi_i\rangle,
\end{equation}
where $c_i^*$ is the complex conjugate of $c_i$.
Note that Eq.~\eqref{eq:sym} specifies the $*$ map only up to a
unitary transformation of no consequence.
One specific realisation of the $*$ map, and what we use in our numerical
examples below, is to first pick a basis $\{|i\rangle\}$ on $\cH$, define
$|\overline i\rangle\equiv |i\rangle$, and then extend the action of $*$ to
arbitrary vectors using the $*$-linearity property.
See also Sec.~3.1 in Ref.~\cite{MUB10} for qubit examples of the $*$ map.

We extend the action of the $*$ map to adjoint vectors,
$*(\langle\psi|)=\langle\overline\psi|=(|\overline\psi\rangle)^\dagger%
=[*(|\psi\rangle)]^\dagger$, and further to the set of operators on $\cH$,
denoted as $\cB(\cH)$,  
\begin{equation}
  *\Bigl(\sum_{ij}c_{ij}|\psi_i\rangle\langle\phi_j|\Bigr)
  \equiv \sum_{ij}c_{ij}^*|\overline\psi_i\rangle\langle\overline\phi_j|.
\end{equation} 
We write $*(X)\equiv \overline X$, for any $X\in\cB(\cH)$.
Note that $\overline{X^\dagger}=(\overline X)^\dagger$, and we denote
$X^\cT\equiv \overline X^\dagger$, a basis-independent transpose operation.
If $X$ is non-negative, then so is $X^\cT$.

Using the $*$ map, we define the \emph{vectorization map}, a \emph{linear} map
from operators to vectors in a vector space $\cV$, $\vc:\cB(\cH)\rightarrow
\cV$,  
\begin{equation}
  \vc(|\psi\rangle\langle \phi|)\equiv*(|\phi\rangle)\otimes |\psi\rangle
  =|\overline \phi\rangle\otimes |\psi\rangle=|\overline \phi\,\psi\rangle,
\end{equation}
for any $|\psi\rangle,|\phi\rangle\in\cH$ and extended to all operators by
linearity.
We write, for any $X\in\cB(H)$, $\vc(X)\equiv |X\rrangle\in\cV$.
Note the useful identity,
\begin{equation}\label{eq:ABCid}
\vc(ABC)=(C^\cT\otimes A)\,\vc(B).
\end{equation}
Also, if $\{|i\rangle\}$ is an orthonormal basis for $\cH$, then so is
$\{|\overline i\rangle\}$. 
Consequently, the vectorized identity operator,
$|\id\rrangle=\vc(\id)=\sum_{i=1}^d|\overline ii\rangle$, can be regarded as a
bipartite maximally entangled (unnormalized) state on $\cH\otimes \cH$. 

Now, we are ready to state the channel-state duality. Consider a CP map,
$\cE:\cB(\cH)\rightarrow \cB(\cH)$, acting as $\cE(\cdot)=\sum_a
E_a(\cdot)E_a^\dagger$ for a (nonunique) set of Kraus operators $\{E_a\}$.
We define
\begin{align}\label{eq:duality}
  \rho_\cE&\equiv \sum_a|E_a\rrangle\llangle E_a|
   =\sum_a(\id\otimes E_a)|\id\rrangle\llangle\id|(\id\otimes E_a^\dagger)
          \nonumber\\
  &=(\id\otimes\cE){\left(|\id\rrangle\llangle\id|\right)},
\end{align}
where we have used the identity in Eq.~\eqref{eq:ABCid};
the $\id$ in $\id\otimes\cE$ denotes the identity map.
Thus defined, $\rho_\cE$ is a nonnegative operator on $\cV$; it can
also be regarded as an unnormalized state (density operator) on the bipartite
Hilbert space $\cH\otimes\cH\equiv \cH_1\otimes\cH_2$, labelling the two
subsystems by $1$ and $2$.
In the latter picture, one regards $|\id\rrangle\llangle\id|$ as the
density operator for a maximally entangled state on $\cH\otimes \cH$, and
$\rho_\cE$ is the density operator that results from the action of the map
$\id\otimes\cE$ on it.  

That $\rho_\cE$ is invariant under a change of Kraus representation for the
$\cE$ is manifest in the last line of Eq.~\eqref{eq:duality}.
We can turn the logic around: Any bipartite state on $\cH\otimes \cH$
possesses a spectral decomposition into eigenvectors, and the identification
of those eigenvectors, with their corresponding (square root of the)
eigenvalues, as vectorized Kraus operators immediately gives an associated CP
map on $\cB(\cH)$.
Equation~\eqref{eq:duality} hence states a duality between CP
maps $\cE$ and states $\rho_\cE\geq 0$. $\rho_\cE$ is sometimes called the
``Choi state'' of the CP map $\cE$.
Observe that
\begin{equation}
  \cE(X)=\tr_{1}\{\rho_\cE(X^\cT \otimes \id)\}.
\end{equation}

We are primarily interested in CP maps that are also TP.
In this case, the state $\rho_\cE$ dual to the CP \emph{and} TP channel
satisfies the partial trace condition, 
\begin{equation}\label{eq:TPCond}
  \tr_2(\rho_\mathcal{E})=\id,
\end{equation}
i.e., $\cE$ is CPTP if and only if $\rho_\cE\geq 0$ and
$\tr_2(\rho_\cE)=\id$.
A simple count verifies that we have just the right number of parameters:
A CP $\cE$ is represented by $d^4$ real parameters---a positivity-preserving
map that specifies how a $d^2$-element basis of operators on $\cH$ is mapped
back to itself---and this is the same number of real parameters needed to
specify an unnormalized non-negative $\rho_\cE$; the TP condition removes $d^2$
parameters, leaving $d^2(d^2-1)$ real parameters for a CPTP map, i.e., a
quantum channel.
Note that the set of $\rho_\cE$s corresponding to quantum channels form a
convex set of states, each with trace $d$. We denote the convex set of all
$\rho_\cE$ that satisfy Eq.~\eqref{eq:TPCond} by $\STP$, and refer to
$\rho_\cE\in\STP$ as a TP state. 

This duality between quantum channels and states enables
us to sample quantum channels with algorithms for sampling
quantum states (see the next section).
Furthermore, the problem of process tomography---the estimation of the full
description of a quantum channel acting on a quantum system---can be re-cast
as that of \emph{state} tomography.
As the applications of our channel sampling algorithm discussed below are
related to estimating quantum channels, we use the remainder of this section
to recall this connection between state and process tomography, stemming from
the channel-state duality \cite{MLEreview}. 

Quantum process tomography seeks to discover the full description of some
unknown quantum channel $\cE$, through $N$ uses of the channel.
Standard strategies involve choosing a set of input states $\{\rho^{(i)}\}$,
sending $N^{(i)}$ copies of state $\rho^{(i)}$ through the channel $\cE$, and
then measuring the output state using a POVM $\Pi^{(i)}\equiv
\{\Pi_k^{(i)}\}$.
For each $i$, the tomographic outcome probabilities come from the Born rule,
\begin{equation}\label{eq:BornRule}
  p_k^{(i)}= \tr\{\Pi_k^{(i)}\cE(\rho^{(i)})\}
  =\tr\Bigl\{\rho_\cE\Lambda_k^{(i)}\Bigr\},
\end{equation} 
where $\Lambda_k^{(i)}\equiv (\rho^{(i)})^\cT\otimes\Pi_k^{(i)}$.
Written in this manner, the expression for $p_k^{(i)}$ reminds one of the
situation of state tomography of $\rho_\cE$, where the set
$\{\Lambda_k^{(i)}\}$ forms a pseudo-POVM in that $\Lambda_k^{(i)}\geq
0\ \forall k,i$, and $\sum_k\Lambda_k^{(i)}=\rho^{(i)}\otimes \id$ for any $i$. 
Note that $\sum_kp_k^{(i)}=1$, as guaranteed by the TP condition in
Eq.~\eqref{eq:TPCond} together with the normalization $\tr(\rho^{(i)})=1$. 

The likelihood function for the data
$D=\{D^{(i)}=(n_1^{(i)},n_2^{(i)},\ldots)\}$---$n_k^{(i)}$ denotes the number
of clicks in detector $\Pi_k^{(i)}$ when $\rho^{(i)}$ is sent, and
${\sum_kn_k^{(i)}=N^{(i)}}$---collected is 
\begin{equation}
  L(D|\rho_\cE)=\prod_iL(D^{(i)}|\rho_\cE)
  =\prod_i{\left[\prod_k(p_k^{(i)})^{n_k^{(i)}}\right]},
\end{equation}
where we omit the combinatorial factors that are needed for proper
normalization but are not important here.
Disregarding quantum constraints, the likelihood is maximized, over all
$\{p_k^{(i)}\}$, by setting ${p_k^{(i)}=\sfrac{n_k^{(i)}}{N^{(i)}}}$; with
quantum constraints, a constrained maximization of $L(D|\rho_\cE)$ over all
permissible probabilities---those $p^{(i)}_k$s that could have
come from a nonnegative $\rho_\cE$ and which satisfy ${\sum_kp_k^{(i)}=1\
 \forall i}$---yields what is known as the maximum-likelihood estimator (MLE)
for $\rho_\cE$~\cite{MLEreview}.

\section{Parameterizing channels}\label{sec:param}
\subsection{Arbitrary channels}\label{sec:GenParam}
To obtain a sample of quantum channels according to some specified
distribution, we generate Choi states $\rho_\mathcal{E}$ with the HMC
algorithm. 
The HMC method demands a parameterization of the state space (in this case the
space of $\rho_\cE$) with no superfluous parameters and no external
constraints.
In Ref.~\cite{Seah+1:15}, the ability to sample quantum states with the HMC
algorithm was demonstrated using a parameterization of the full quantum state
space.
Because of the TP condition, sampling of quantum channels demands a
parameterization of, not the full quantum state space as in
Ref.~\cite{Seah+1:15}, but only of the set $\STP$ of TP states.
Here, as our central result, we explain how to accomplish this.

We first choose a product basis $\{|\overline ij\rangle\}_{i,j=1}^d$ on
$\cH\otimes\cH$ and represent $\rho_\cE$ as a $d^2\times d^2$ matrix---also 
denoted as $\rho_\cE$, to simplify notation---with complex entries.
Positivity of $\rho_\cE$ means that we can write $\rho_\cE=A^\dagger A$, where
$A$ is a $d^2\times d^2$ upper triangular complex matrix with real entries in
the last column.
The $d^2$ columns of $A$ are labelled using a double index,
\begin{equation} \label{eq:AMatrix}
A = \begin{pmatrix}
| & | &  & | \\
\varphi_{11}     & \varphi_{12} & \dots  & \varphi_{dd} \\
| & | & & |
\end{pmatrix} ,
\end{equation}
so that $\rho_\cE=\sum_{ijkl=1}^d\varphi_{ij}^\dagger\varphi_{kl}|\overline
ij\rangle\langle \overline kl|$, as the abstract, basis-independent object.
Stacking the columns of $A$ to form columns with $d^3$ entries,
\begin{equation}
\varphi_i\equiv \begin{pmatrix}
\varphi_{i1} \\
\varphi_{i2} \\
\vdots \\
\varphi_{id}
\end{pmatrix},\quad \textrm{for }i=1, 2,\ldots, d,
\end{equation}
permits writing the TP condition in Eq.~\eqref{eq:TPCond}, that is
${\tr_2(\rho_\cE)=\sum_{ij}{\left(\sum_k\varphi_{ik}^\dagger\varphi_{jk}\right)}%
  |\overline i\rangle\langle\overline j|=\id
  =\sum_{ij}\delta_{ij}|\overline i\rangle\langle\overline j|}$, as 
an orthonormality condition on the $\varphi_i$s, 
\begin{equation}
\varphi_i^{\dagger}\varphi_j = \delta_{ij} \quad\textrm{for }i,j =1,2,\ldots, d.
\end{equation}
Hence, to sample quantum channels, we simply need to find a parameterization
for the orthonormal set $\{\varphi_i\}_{i=1}^d$.  

Let us count the number of parameters needed.
Since $A$ is upper triangular, $\varphi_{ik}$ has $(ik)+1$ generically
nonzero entries, where $(ik)\equiv (i-1)d+(k-1)$ is a $d$-nary number.
Each $\varphi_i$ thus has $K_i\equiv \sum_k[(ik)+1]=id^2-\frac{1}{2}d(d-1)$
nonzero entries.
These nonzero entries are all complex, except for the $d^2$ of them in
$\varphi_{dd}$, which are real.
The orthonormality conditions on the $\varphi_i$s remove $d^2$ real
parameters.
Altogether then, the $\varphi_i$s are described by
$2\sum_iK_i-d^2-d^2=d^2(d^2-1)$ real parameters, exactly the number needed to
describe a quantum channel. 

To specify an appropriate parametrization of the $\varphi_i$ set, it is
convenient to reshuffle the rows of $\varphi_i$ so that all the
identically-zero entries of each $\varphi_i$ are collected together.
We first define the matrix
\begin{equation}
\Phi\equiv \begin{pmatrix}
| & | &  & | \\
\varphi_1     & \varphi_2 & \dots  & \varphi_d \\
| & | & & |
\end{pmatrix}.
\end{equation}
Observe that the orthonormality conditions on the $\varphi_i$s translate into
the requirement that $\Phi^\dagger \Phi=\id$. 
Let $P$ be a $d^3\times d^3$ permutation matrix such that
\begin{equation}
\Psi\equiv P\Phi=\begin{pmatrix}
| & | &  & | \\
\psi_1     & \psi_2 & \dots  & \psi_d \\
| & | & & |
\end{pmatrix}
\end{equation}
has columns $\psi_i$s, each of which is a reshuffled $\varphi_i$ with
all identically zero entries located below the generically nonzero ones, i.e.,
the $k$th entry of $\psi_i$, which we denote as $\psi_{ik}$, is generally
nonzero for $k=1,\ldots,K_i$, and zero for $k=K_i+1,\ldots, d^3$.
Such a $P$ matrix exists because $A$ is upper triangular.
Requiring $\Phi^\dagger \Phi=\id$ is equivalent to demanding
$\Psi^\dagger\Psi=\Phi^\dagger P^{-1}P\Phi=\id$.  

We are now ready to state the parameterization for the $\psi_i$s, thereby
giving a parameterization for $\STP$.
We begin with $\psi_d$, parameterizing it with spherical coordinates so that
it is normalized,  
\begin{align}
\psi_{dk}\equiv{\left\{\begin{array}{ll}
\upe^{\upi\phi_k}(\cos\theta_{k-1})S_k&\textrm{for }k=1,\ldots,K_d\\
0&\textrm{for }k=K_d+1,\ldots,d^3
\end{array}\right.},
\end{align}
where $\theta_0\equiv 0$ fixed, and the $S_k$s are recursively defined
as $S_k=(\sin\theta_k) S_{k+1}$, with $S_{K_d}=1$.
Here, the $\phi_k$s for the $\psi_{dk}$s that come from the real entries of
$\varphi_{dd}$ are understood to be set to zero (which ones they are,
depends on the choice of $P$).  
$\psi_d$ is hence parameterized by real parameters $\theta_1,\ldots,
\theta_{K_d-1}$, and $K_d-d^2$ $\phi_k$ (real) parameters, giving $2K_d-1-d^2$
real parameters in all.
Note the identity, 
\begin{equation}
\sum_{k=1}^m|\psi_{dk}|^2=S_m^2, \quad\textrm{for any }m=1,2,\ldots, K_d,
\end{equation}
so that the norm-square of $\psi_d$ is simply
$\psi_d^\dagger\psi_d=\sum_{k=1}^{K_d}|\psi_{dk}|^2=S_{K_d}^2=1$, i.e.,
$\psi_d$ has length~1. 

Next, let $v_n$, for $n=1, \ldots, K_{d-1}-1$, be the $d^3$-long column vector
with the $k$th entry defined as 
\begin{align}
v_{nk}&\equiv \sfrac{1}{S_{n+1}}{\left\{\begin{array}{ll}
   \psi_{dk}\big\vert_{\theta_n\rightarrow\theta_n+\frac{\pi}{2}}
  &\textrm{for }k=1,\ldots,n+1\\[1ex]
 0&\textrm{for }k=n+2,\ldots,d^3
\end{array}
\right.}\\
&=\sfrac{1}{S_{n+1}}{\left\{\begin{array}{ll}
\psi_{dk}\sfrac{\cos\theta_n}{\sin\theta_n}&\textrm{for }k=1,\ldots,n\\[1ex]
\psi_{d(n+1)}\sfrac{-\sin\theta_n}{\cos\theta_n}&\textrm{for }k=n+1\\[1ex]
0&\textrm{for }k=n+2,\ldots,d^3
\end{array}
\right.}.\nonumber
\end{align}
Observe that $v_n$ is orthogonal to $\psi_d$, for every $n$, since
\begin{align}
  S_{n+1}v_n^\dagger \psi_d
  &=\sfrac{\cos\theta_n}{\sin\theta_n}
    \sum_{k=1}^n|\psi_{dk}|^2-\sfrac{\sin\theta_n}
    {\cos\theta_n}|\psi_{d(n+1)}|^2\\ 
  &=\cos\theta_n\sin\theta_nS_{n+1}^2-\sin\theta_n\cos\theta_nS_{n+1}^2=0
    \nonumber.
\end{align}
One can check, in a similar manner, that the $v_n$ column vectors form an
orthonormal set.  

The span of $\{v_n\}_{n=1}^{K_{d-1}-1}$ lies in the orthogonal subspace of
$\psi_d$.
$\psi_1,\psi_2,\ldots\psi_{d-1}$ are to be orthogonal to $\psi_d$, so we can
set them to be in the linear span of $\{v_n\}$.
Note the both $\psi_{d-1}$ and $v_{K_{d-1}-1}$ have the same number ($=K_{d-1}$)
of nonzero entries, the largest among the $\psi_i$s ($i=1,\ldots,{d-1}$) and
$v_n$s.
Specifically, we define
\begin{equation}
\begin{pmatrix}
| & | &  & | \\
\psi_1     & \psi_2 & ...  & \psi_{d-1} \\
| & | & & |
\end{pmatrix}\equiv V\widetilde\Psi,
\end{equation}
where $V$ is the (non-square) matrix with columns $v_1, v_2,\ldots,
v_{K_{d-1}-1}$.
$\widetilde\Psi$ is defined such that its columns are the coefficients of the
$\psi_i$s when expressed as a linear combination of the $v_n$s, i.e.,
$\psi_i=V\widetilde\psi_i=\sum_n\widetilde\psi_{in}v_n$, where
$\widetilde\psi_i$ is the $i$th column of $\widetilde\Psi$, and
$\widetilde\psi_{in}$ are its entries.
Note that $V$ is a $d^3\times (K_{d-1}-1)$ matrix with the last $d^3-K_{d-1}$
rows completely zero, while $\widetilde\Psi$ is a $(K_{d-1}-1)\times (d-1)$
matrix. 

Observe that the orthonormality of the $\psi_i$s, for
$i=1,\ldots,{d-1}$ is equivalent to the orthonormality of the columns of
$\widetilde\Psi$, i.e., $\widetilde\Psi^\dagger \widetilde\Psi=\id$.
This is then the same problem as before, for $\Psi$, with now one fewer
column.
We hence repeat the procedure above, parameterizing $\widetilde\psi_{d-1}$
using a new set of spherical coordinates ($\theta$s and $\phi$s; note that
none of the $\phi$s are set to zero as the $\psi_{i\neq d}$s are
generally complex), defining new $v$ vectors orthogonal to it, getting a new 
$\widetilde\Psi$, and so forth.
We do this recursively until all $\psi_i$s are parameterized.

Let us check that the recursive procedure yields the right number of
parameters for the full set of orthonormal $\psi_i$s.
As mentioned earlier, in the first round, $\psi_d$ (and the $V$ there) is
parameterized by $2K_d-1-d^2$ parameters, that subtraction of $d^2$ coming
from the $d^2$ zero $\phi_k$s done for $\psi_d$ only.
In the next round, $\psi_{d-1}$ is parameterized by an additional (on top of
the ones that go into $V$) $2(K_{d-1}-1)-1$ real parameters; in yet the next
round, $\psi_{d-2}$ is parameterized by an additional $2(K_{d-2}-2)-1$ real
parameters; and so forth.
Altogether then, we have
$-d^2+\sum_{i=0}^{d-1}{\left[2(K_{d-i}-i)-1\right]}=d^4-d^2$ real parameters,
exactly the right number needed for parameterizing $d$-dimensional quantum
channels. 

To illustrate how one applies the above parameterization, the case of qutrit
channels is discussed in Appendix~\ref{app:Qutrit}.
In the following sections, we make use of our parameterization in a HMC
algorithm to sample quantum channels according to specified distributions, and
demonstrate the usefulness of these samples in different applications.
Before we get to that, however, let us mention a parameterization designed
specifically for unital qubit channels, useful for one of our examples below.

\subsection{Unital qubit channels} \label{sec:UnitalParam}
A useful class of quantum channels is the set of unital channels,
those that preserve the identity operator, $\cE(\id)=\id$.
The unitality condition can be stated in terms of the Choi state as the
requirement 
\begin{equation}\label{eq:unital}
  \tr_1(\rho_\cE)=\id.
\end{equation}
A unital quantum channel thus has $\rho_\cE$ such that $\tr_i(\rho_\cE)=\id$
for $i=1,2$, stating both the TP and unitality conditions.
This is generally a difficult pair of conditions to impose, for a
parameterization of unital channels with exactly the right number of
parameters, as needed for HMC. 

For unital qubit channels, however, this can be done in a straightforward
manner, as we describe here \cite{MagicBasis}. 
The Choi state of a qubit channel is a two-qubit state.
Any two-qubit state (normalized to trace 2) can be written as
\begin{equation}
  \rho = \tfrac{1}{2}(\id + \bm{\sigma}\cdot\bm{s}
         + \bm{t}\cdot\bm{\tau} + \bm{\sigma}\cdot \bm{C}\cdot\bm{\tau}),
\end{equation}
where $\bm{\sigma} = (\sigma_x,\sigma_y,\sigma_z)$ is the vector of Pauli
operators for the first qubit and $\bm{\tau} = (\tau_x,\tau_y,\tau_z)$ is the
vector of Pauli operators for the second qubit.
(Here, the word ``vector'' is used in the physicist's sense of a
three-dimensional spatial vector.)
$\bm{s}$ and $\bm{t}$ are the Bloch vectors for qubits 1 and 2, respectively;
$\bm{C}$ is a dyadic, representable by a $3\times 3$ matrix of real numbers
corresponding to the coefficients of $\sigma_i\tau_j$, for $i,j=x,y,z$. 
The TP condition requires $\bm{s}=0$; the unitality condition demands
$\bm{t}=0$.
The Choi state of a unital qubit channel thus takes the form
\begin{equation}
  \rho_\cE = \tfrac{1}{2}(\id + \bm{\sigma}\cdot \bm{C}\cdot\bm{\tau}).
\end{equation}

Up to local unitary transformation, the dyadic $\bm{C}$ can always be chosen
to be diagonal $\bm{C}_{\text{diag}}$. 
For $\rho_\mathcal{E}$ to be positive semi-definite, the three diagonal
entries of $\bm{C}_{\text{diag}}$ must lie within a tetrahedron with the
vertices  
\begin{align}\label{eq:tetra}
  &\bm{v}_1 = (-1,-1,-1),\nonumber\\
  &\bm{v}_2 = (-1,1,1), \nonumber\\
  &\bm{v}_3 = (1,-1,1), \nonumber\\
  \textrm{and}\quad&\bm{v}_4 = (1,1,-1),
\end{align}
where each vertex corresponds to one of four pairwise orthogonal
maximally entangled two-qubit states.
We parameterize the three entries of $\bm{C}_{\text{diag}}$ by the convex
combination of the four vertices 
\begin{equation}
  (c_1,c_2,c_3) = \alpha_1 \bm{v}_1 + \alpha_2 \bm{v}_2
                 + \alpha_3 \bm{v}_3 + \alpha_4 \bm{v}_4,
\end{equation}
where
\begin{align}
  &\alpha_1 = \cos^2\theta_1 \nonumber\\
  &\alpha_2 = \sin^2\theta_1\cos^2\theta_2 \nonumber\\
  &\alpha_3 = \sin^2\theta_1\sin^2\theta_2\cos^2\theta_3 \nonumber\\
  &\alpha_4 = \sin^2\theta_1\sin^2\theta_2\sin^2\theta_3.
\end{align}
Generally, the dyadic $\bm{C}$ can be written as
\begin{equation} \label{eq:dyadic}
  \bm{C} = \bm{R_1} \bm{C}_{\text{diag}} \bm{R_2}^T
\end{equation}
where $\bm{R_1}$ and $\bm{R_2}$ are the rotation matrices representing the
local unitary transformations (equivalently, spatial rotations in the
Bloch-ball picture) of qubits 1 and 2, respectively.
$\bm{R_1}$ and $\bm{R_2}$ can each be parameterized by three rotation angles.
Altogether, we have a parameterization of the set of all unital qubit
channels, specified by nine angle parameters.

\section{Applications}\label{sec:applications}
The HMC algorithm is a method for generating random samples from any target distribution by making use of pseudo-Hamiltonian dynamics in a mock phase space. Upon identifying the parameters in the parameterizations given in Sec.~\ref{sec:param}---which satisfy the requirements of not having superfluous parameters and no constraints---as the position variables in the mock phase space, we can employ the HMC algorithm. We give a brief review of the HMC algorithm in Appendix~\ref{app:HMC}. A more detailed discussion can be found in Ref.~\cite{Seah+1:15}. In this section, we demonstrate the use of random samples of channels in three applications related to process tomography.
That the examples are related to tomography simply reflects the authors'
original motivation and source of interest in the matter of channel sampling.
The channel parameterization invented here and the resulting ability to sample
according to a user-specified distribution using a HMC algorithm are
applicable beyond tomography tasks.

\subsection{Error regions for process estimation}\label{sec:OER}
Whether one chooses to use the MLE or some other estimator for $\rho_\cE$, the
\emph{point} estimator will not coincide exactly with the true $\rho_\cE$ with
finite data.
It is important then to endow the point estimators with error regions
expressing the uncertainty in our knowledge of the identity of the channel.
Here, we adopt as error regions the notion of smallest credible regions (SCRs)
proposed in Ref.~\cite{OER13}.
SCRs were originally proposed for the estimation of quantum states,
whether they are TP states or not, but completely analogous notions can be
defined for $\STP$.  
Here, we examine the construction of SCRs for
the task of quantum process estimation, as an application of our channel
sampling algorithm.
We first recall a few key points about SCRs pertinent to our discussion here;
the reader is referred to \cite{OER13} for further details. 

The SCR is the region---a set of states---in $\STP$ with the smallest size for
a chosen credibility. Size is the prior content of a region in $\STP$, i.e.,
the prior (before any data are taken) probability that the true state is in
the region; credibility is the posterior (after incorporating the data)
content of that region.
The SCRs are bounded-likelihood regions (BLRs), i.e., regions
$\cR_\lambda$ comprising all states with likelihood no smaller than a
threshold fraction $\lambda\in[0,1]$ of the maximum likelihood
$L_{\max}(D)$,  
\begin{equation}
\cR_\lambda(D)=\{\rho\in\STP:L(D|\rho)\geq\lambda L_{\max}(D)\},
\end{equation}
with $\cR_0=\STP$.
The size $s_\lambda$ of the BLR $\cR_\lambda$ is its prior
content, and its credibility $c_\lambda$ is its posterior content, 
\begin{equation} \label{eq:slambda}
  s_\lambda(D)=\!\int\limits_{\cR_\lambda(D)}\!\!(\upd\rho)
  \quad \textrm{and}\quad
  c_\lambda(D)=\!\int\limits_{\cR_\lambda(D)}\!\!(\upd\rho)
  \,\frac{L(D|\rho)}{L(D)},
\end{equation}
with ${s_0=c_0=1}$ when ${\lambda=0}$.
The volume element $(\upd\rho)$ expresses the prior distribution;
$(\upd\rho)\sfrac{L(D|\rho)}{L(D)}$ is the posterior distribution.
$L(D)\equiv \int_{\cR_0}(\upd \rho)L(D|\rho)$, a normalizing factor, is the
likelihood of obtaining the data $D$ for the chosen prior.  
For tomography problems, it is often natural to state the prior distribution
in terms of the POVM-induced probabilities [see Eq.~\eqref{eq:BornRule}], 
\begin{equation}
(\upd\rho) = (\upd p)\,w_0(p),
\end{equation}
where $w_0(p)$ is the prior density, nonzero only for
$p\equiv(p_1^{(1)},p_1^{(2)},\ldots,p_2^{(1)},\ldots)$ that corresponds to a
$\rho\in\STP$, and $(\upd p)\equiv\upd p_1^{(1)}\upd p_1^{(2)}\cdots~$. 

To report the error region for an experiment with data $D$, following the
scheme of Ref.~\cite{OER13}, $s_\lambda$ and $c_\lambda$ are calculated
for all values of $\lambda$.
The error regions are reported by plotting $s_\lambda$ and $c_\lambda$ as
functions of $\lambda$.
For a desired level of credibility, the $\lambda$ value is read off, and the
error region is the $\cR_\lambda$ for that value of $\lambda$.  
The size and credibility of a BLR [see Eq.~\eqref{eq:slambda}] cannot,
in general, be computed analytically, due to the complicated integration
region.
Instead, we make use of MC integration:
We generate random samples using HMC according to the prior and posterior
distributions; the size and credibility are then the fractions of points
contained in the BLR for the two distributions. 

A related concept is the plausible region \cite{Evans:15}.
This is the set of all points in $\STP$, for which the data provide evidence
in favor of---${L(D|\rho)>L(D)}$.
The plausible region is in fact a BLR, with a critical value of $\lambda$,
\begin{equation}
\lambda_{\text{crit}}(D) = \frac{L(D)}{L_{\text{max}}(D)}. 
\end{equation}
Once we have computed the size and credibility curves, we can also identify
the plausible region for the data. 

\FigOne

As a first example, we look at single-qubit channels.
The input states $\rho^{(i)}$ for process tomography are taken to be the
tetrahedron states, 
\begin{equation}\label{eq:TetraStates}
  \rho^{(i)}
  = \tfrac{1}{2}(\id+\bm{a}_i\cdot\bm{\sigma}),\quad  i=1,2,3,\textrm{ and }4,
\end{equation}
where $\bm{a}_i=3^{-1/2}\bm{v}_i$ with the vertex vectors of
Eq.~\eqref{eq:tetra}.  
For every $i$, we use the same POVM, the four-outcome tetrahedron measurement,
with outcomes 
\begin{equation}\label{eq:TetraPOVM}
  \Pi_k = \tfrac{1}{4}(\id+\bm{a}_k\cdot\bm{\sigma}),
  \quad  k=1,2,3,\textrm{ and }4.
\end{equation}
We simulate data using an amplitude-damping channel described by the Kraus
operators 
\begin{equation}
  E_0\equiv {\left(\begin{array}{cc}
                   1&0\\0&\sqrt{1-\gamma}
                   \end{array}\right)}\quad \textrm{and}\quad 
  E_1\equiv {\left(\begin{array}{cc}
                    0&\sqrt{\gamma}\\0&0
                    \end{array}\right)},
\end{equation}
where $\gamma$, the damping parameter, is set to $0.4$.
The matrices above refer to the computational basis. For the matrices appear in the rest of this paper, it should be assumed that they refer to the computational basis as well.
24 copies of each input state $\rho^{(i)}$ are measured (simulated), giving a
total of 96 counts over the four input states. The simulated data are reported in Table~\ref{tab:Data1}.
\begin{table}[]
\caption{Simulated data for the first example in Sec.~\ref{sec:OER}. Each entry $n_k^{(i)}$ represents the number of clicks in detector $\Pi_k$ when the input state $\rho^{(i)}$ is sent.}
\label{tab:Data1}
\begin{tabular}{cc}
\raisebox{-0.4cm}{$\rho^{(i)}$} &
\begin{tabular}{cc|cccc}
\multicolumn{6}{c}{\hspace{0.6cm}$\Pi_k$}\\
  &  & 1 & 2 & 3 & 4 \\\hline
1 &  & 7 & 5 & 6 & 6 \\
2 &  & 3 & 13 & 4 & 4 \\
3 &  & 2 & 7 & 8 & 7 \\
4 &  & 3 & 7 & 8 & 6
\end{tabular}
\end{tabular}
\end{table}

For the prior distribution, we choose the conjugate prior,
\begin{equation}
  (\upd p)\,w_0(p)\propto(\upd p)
  \prod_{i,k=1}^4 \bigl(p_k^{(i)}\bigr)^{48\bar{p}_k^{(i)}},
\end{equation}
where ${\bar{p}=\{\bar{p}_k^{(i)}\}}$ corresponds to the Born probabilities 
[see Eq.~\eqref{eq:BornRule}] for an amplitude-damping channel with
$\gamma=0.5$, expressing our prior belief that that is the actual channel.
Figure \ref{fig:SCR}(a) shows the size and credibility curves, obtained from
MC integration using 500,000 sample points generated from HMC with the channel
parameterization of Sec.~\ref{sec:param}.
The critical $\lambda$ value for the plausible region is indicated with a red
dashed line, with size value $s=0.2102$ and credibility value $c=0.8586$.
The true channel is contained in all BLRs with $\lambda<0.0302$ and
$c_\lambda>0.5511$, and is thus in the plausible region. 

Now, qubit channels are simple to characterize and there are many ways of
sampling from the space of qubit channels.
It is hence useful to see how our sampling algorithm works for examples beyond
the qubit situation, for which proper sampling is more challenging.
As a second example, we consider an amplitude-damping qutrit
(three-dimensional quantum system) channel with the Kraus operators 
\begin{align}
&E_0\equiv {\left(\begin{array}{ccc}
1&0&0\\0&\sqrt{1-\gamma_1}&0\\0&0&\sqrt{1-\gamma_2}
\end{array}\right)}, \\
&E_1\equiv {\left(\begin{array}{ccc}
0&\sqrt{\gamma_1}&0\\0&0&0\\0&0&0
\end{array}\right)}, \quad
\textrm{and}\quad E_2\equiv {\left(\begin{array}{ccc}
0&0&\sqrt{\gamma_2}\\0&0&0\\0&0&0
\end{array}\right)},\nonumber
\end{align}
for $\gamma_1=0.1$ and $\gamma_2=0.5$. 

The POVM used is one of the symmetric, informationally complete POVM
(SIC-POVM) from the one-parameter family of qutrit SIC-POVMs.
It can be described by a set of states $\{\ket{\mu_i}\}$;
when written in the computational basis, they are given explicitly by  
\begin{align}
&\begin{pmatrix}
\ket{\mu_1} \ket{\mu_2} \cdots \ket{\mu_9}
\end{pmatrix} \nonumber\\
&\equiv\frac{1}{\sqrt{2}}\begin{pmatrix}
1 & 1 & 1 & 0 & 0 & 0 & \omega & \omega^{*} & 1 \\
\omega & \omega^{*} & 1 & 1 & 1 & 1 & 0 & 0 & 0 \\
0 & 0 & 0 & \omega & \omega^{*} & 1 & 1 & 1 & 1 
\end{pmatrix},
\end{align}
where $\omega =\Expi{2\pi/3}$, $\omega^*=\omega^2$, and $1+\omega+\omega^2=0$.
The POVM elements are
\begin{equation}
\Pi_i = \frac{1}{3}\ket{\mu_i}\bra{\mu_i},\quad  i=1,2,\cdots,9.
\end{equation}
The input states are
\begin{equation}
\rho^{(i)} = \ket{\mu_i}\bra{\mu_i},\quad  i=1,2,\cdots,9.
\end{equation}

For each of the input states, the number of copies measured is 27, giving a
total of 243 counts. The simulated data are reported in Table~\ref{tab:Data2}.
\begin{table}[]
\caption{Simulated data for the second example in Sec.~\ref{sec:OER}. Each entry $n_k^{(i)}$ represents the number of clicks in detector $\Pi_k$ when the input state $\rho^{(i)}$ is sent.}
\label{tab:Data2}
\begin{tabular}{cc}
\raisebox{-0.6cm}{$\rho^{(i)}$} &
\begin{tabular}{cc|ccccccccc}
\multicolumn{11}{c}{\hspace{0.7cm}$\Pi_k$}\\
  &  & 1 & 2 & 3 & 4 & 5 & 6 & 7 & 8 & 9 \\\hline
1 &  & 8 & 1 & 4 & 1 & 8 & 2 & 1 & 2 & 0 \\
2 &  & 3 & 10 & 3 & 1 & 1 & 2 & 3 & 3 & 1 \\
3 &  & 1 & 1 & 9 & 1 & 2 & 2 & 3 & 5 & 3 \\
4 &  & 8 & 3 & 3 & 4 & 0 & 1 & 4 & 3 & 1 \\
5 &  & 2 & 3 & 3 & 1 & 10 & 2 & 4 & 2 & 0 \\
6 &  & 3 & 4 & 4 & 2 & 0 & 8 & 2 & 2 & 2 \\
7 &  & 3 & 2 & 4 & 2 & 0 & 2 & 9 & 4 & 1 \\
8 &  & 2 & 0 & 4 & 0 & 0 & 2 & 6 & 8 & 5 \\
9 &  & 3 & 3 & 3 & 1 & 5 & 0 & 1 & 3 & 8 \\
\end{tabular}
\end{tabular}
\end{table}
The prior is the primitive prior, i.e., $w_0(p)$ is a
constant wherever it is nonzero.
Figure \ref{fig:SCR}(b) shows the size and credibility curves, obtained from
MC integration with 100,000 sample points using HMC and our channel
parameterization.
As before, the critical $\lambda$ value for the plausible region is indicated
by the vertical dashed line.
The size and credibility of the plausible region are $s=0.0032$ and $c=0.9990$
respectively.
The true channel is contained in all BLRs with $\lambda<1.1560\times10^{-6}$,
and is thus in the plausible region.

\subsection{Marginal likelihood for channel properties}\label{sec:OEI}
Often, one is only interested in certain properties of a channel, like the
fidelity between the output of the channel and its input, rather than a full
channel description in the form of its process matrix. 
If one could directly measure that one quantity of interest, one expects to
accomplish the estimation task with significantly fewer uses of the channel
than needed for full tomography.
However, a direct measurement of the quantity of interest may be difficult to
design and implement, while the process tomography measurement is often
standard procedure.
Even in the latter case, one should still estimate the quantity of interest
directly from the tomography data, rather than first estimating the full
process matrix and then computing the quantity of interest from that estimate
\cite{OEI16}. 

The key ingredient in making inferences about a property $F$ of a channel from
tomographic data $D$ is the marginal likelihood, obtained by integrating the
full likelihood $L(D|p)$ over the irrelevant parameters, 
\begin{eqnarray} \label{eq:marlik1}
  L(D|F)&=& \frac{\int(\mathrm{d}p)\,w_r(p)\,\delta\bigl(F-f(p)\bigr)\,L(D|p)}
                {\int(\mathrm{d}p)\,w_r(p)\,\delta\bigl(F-f(p)\bigr)}
  \nonumber \\ &\equiv& \frac{W_{r,D}(F)}{W_{r,0}(F)},
\end{eqnarray} 
where $W_{r,D(0)}(F)$ is the integral in the numerator(de\-nominator).
$f(p)$ is the function that expresses $F$ in terms of the tomographic
probabilities $p$, and $w_r(p)$ is the prior density on $p$, which induces a
prior density on~$F$.
$\delta\bigl(F-f(p)\bigr)$ is the Dirac delta function that enforces
${f(p)=F}$. 
Once we have the marginal likelihood, we can proceed in an analogous way as in
Sec.~\ref{sec:OER} to construct the smallest credible interval (SCI) and the
plausible interval for $F$, as well as perform other statistical inference
tasks based on the marginal likelihood. 

We thus need a general procedure for computing the marginal likelihood
$L(D|F)$. 
In Ref.~\cite{OEI16}, an iterative algorithm was developed for that
purpose, requiring the use of random samples according to specified
distributions.
The reader is referred to Ref.~\cite{OEI16} for the full description of
the iterative algorithm, and to Appendix \ref{app:Iter} for the details
relevant for our examples below.
Here, we give only a brief account of the basic ideas.
The delta functions in the defining equation \eqref{eq:marlik1} are
difficult to handle in a numerical evaluation of the integrals.
Instead, we evaluate the antiderivatives $P_{r,i}(F)$, with respect to $F$, of
$W_{r,i}(F)$, 
\begin{equation}
    P_{r,i}(F)\equiv \int\upd F\,W_{r,i}(F),\quad i=D,0,
\end{equation}
with step functions in place of the delta functions.
$P_{r,i}$ can be computed by MC integration.
The results are closely fitted with several-parameter functions, and then
differentiated to give $W_{r,i}$, and hence the marginal likelihood.  
This procedure works, in principle; in practice, one runs into numerical
accuracy problems.
If $w_r(p)$ has little weight over some range of $F$, a rather generic
situation, $P_{r,0}$ will be very flat there, and its derivative cannot be
reliably estimated.  
To overcome this problem, the crux is to note that, because of the delta
functions, the marginal likelihood is invariant under the replacement
$w_r(p)\rightarrow w_r(p)g(f(p))$ for any function $g(F)$ positive over the
entire range of $F$.
We thus have the freedom to choose the $w_r(p)$ used to evaluate $L(D|F)$.
This freedom of choice is exploited in the iterative procedure described in
Ref.~\cite{OEI16}, where the estimate of $W_{r,0}$ is successively
improved by using an $w_r(p)$ modified by the previous (possibly inaccurate)
estimate of $W_{r,0}$, until the desired convergence level is reached.
Each iterative step requires the ability to sample according to the new
$w_r(p)$; that is where the HMC algorithm, permitting sampling in accordance
to a user-specified distribution, comes in.

Below, we carry out the iterative algorithm and compute the marginal
likelihood for two common channel properties, average fidelity $\Favg$ and
minimum fidelity $\Fmin$.
We make use of the HMC algorithm made possible by our channel parameterization
of Sec.~\ref{sec:param}.
Both examples are for qubit channels, and use the same (simulated) tomographic
data obtained from tetrahedron input states [see
Eq.~\eqref{eq:TetraStates}] and the tetrahedron POVM [see
Eq.~\eqref{eq:TetraPOVM}] for the true channel 
\begin{equation}\label{eq:PauliNoise}
  \cE_\mathrm{Pauli}(\cdot)
  \equiv \Bigl(1-\sum_{i=x,y,z}p_i\Bigr)(\cdot)
  +\sum_{i=x,y,z}p_i\sigma_i(\cdot)\sigma_i,
\end{equation}
a Pauli channel.
Here, the $\sigma_i$s are the standard Pauli operators, and
$(p_x,p_y,p_z)=(0.05,0.15,0.2)$.
The data are generated from 96 uses of the channel. The simulated data are reported in Table~\ref{tab:Data3}.
\begin{table}[]
\caption{Simulated data for the examples in Sec.~\ref{sec:OEI}. Each entry $n_k^{(i)}$ represents the number of clicks in detector $\Pi_k$ when the input state $\rho^{(i)}$ is sent.}
\label{tab:Data3}
\begin{tabular}{cc}
\raisebox{-0.4cm}{$\rho^{(i)}$} &
\begin{tabular}{cc|cccc}
\multicolumn{6}{c}{\hspace{0.6cm}$\Pi_k$}\\
  &  & 1 & 2 & 3 & 4 \\\hline
1 &  & 9 & 4 & 4 & 7 \\
2 &  & 6 & 6 & 3 & 9 \\
3 &  & 3 & 5 & 10 & 6 \\
4 &  & 8 & 4 & 5 & 7
\end{tabular}
\end{tabular}
\end{table}
We regard the Pauli channel as noise acting on our quantum system.
We are interested in the fidelity measures, $\Favg$ and $\Fmin$, quantifying
the effect of this noise channel on our system.

\subsubsection{Average Fidelity}
The average fidelity $\Favg$ is defined here as the \mbox{(squared-)fidelity}
between the input and output of the channel $\cE$, averaged over all input
pure states according to the Haar measure.  
We write $F(\psi,\rho)\equiv \langle\psi|\rho|\psi\rangle$ for the square of
the fidelity between a pure state $\psi\equiv|\psi\rangle\langle\psi|$ and an
arbitrary state $\rho$.
Then, the average fidelity for the channel $\cE$ is
\begin{align}
  \Favg(\cE)
  &\equiv \int \upd\psi \,\langle\psi|\cE(\psi)|\psi\rangle \nonumber\\
  &= \langle\psi_0{\left[\int \upd U \,U^\dagger\cE (U \psi_0 U^\dagger)
       U\right]}|\psi_0\rangle \nonumber\\
  &= \frac{1}{d}{\left[1+(d-1)q\right]}.
\end{align}
Here, $\upd U$ is the Haar measure for the space of unitary operators, and
$\psi_0$ is some fiducial pure state.  
In arriving at the last line, we have used a standard result of the twirling
operation \cite{Emerson05} (namely, the expression in the brackets in the
second-to-last line), with $q$ given by  
\begin{equation}\label{eq:q-def}
  q\equiv \frac{1}{d^2-1}\sum_i\tr{\left(\rho_\cE(O_i^\cT\otimes O_i)\right)},
\end{equation}
where $O_i$s are all the traceless elements of an orthonormal (according to
the Hilbert-Schmidt inner product) operator basis, containing an element
proportional to the identity operator, for the $d$-dimensional $\cH$.
In the qubit case, $q$ has the explicit formula,
\begin{equation}\label{eq:q-qubit}
  q = \frac{1}{3}\tr\bigl(\rho_\mathcal{E}(\sigma_x\otimes\sigma_x
-\sigma_y\otimes\sigma_y+\sigma_z\otimes\sigma_z)\bigr),
\end{equation}
where we have chosen the $*$ map such that $|\overline i\rangle=|i\rangle$ for
$\{|i\rangle\}_{i=0}^1$, the $\sigma_z$-basis for the qubit (see comment about
this choice in the second paragraph of Sec.~\ref{sec:isomorphism}). 

\FigTwo\FigThree

We use the iterative procedure of Ref.~\cite{OEI16} to compute the
marginal likelihood $L(D|\Favg)$, for $F\equiv\Favg$.
The final result is shown in Fig.~\ref{fig:avgMarginalLikelihood}; the
intermediate steps of the iterative algorithm are described in Appendix
\ref{app:Supp1}.
With the marginal likelihood at hand, as an example of its usefulness, we can
construct, as in Sec.~\ref{sec:OER}, the SCI for our estimate of $\Favg$.
Figure \ref{fig:OEI}(a) gives the size and credibility curves, as well as the
critical $\lambda$ value for the plausible region. 
Figure \ref{fig:OEI}(b) shows the SCI for $\Favg$ for different
credibility values.
The horizontal black line specifies the plausible interval, which includes the
true value of ${\Favg=0.7333}$ (indicated with an arrow).

\subsubsection{Minimum fidelity of unital qubit channels}
As a second example, also to illustrate the use of the parameterization of
the unital qubit channels of Sec.~\ref{sec:UnitalParam}, we look at the
minimum, or worst-case, (squared-)fidelity of a unital channel.
The minimum fidelity for a channel $\cE$ is the fidelity of the output of
$\cE$ with its (pure) input, minimized over all input states, i.e., 
\begin{equation}
  \Fmin \equiv \min_{\ket\psi} F\bigl(\psi,\cE(\psi)\bigr).
\end{equation}
In the qubit case, $\Fmin$ can be written explicitly using the Bloch-ball
representation as 
\begin{equation}
  \Fmin = \min_{\bm{s}: |\bm{s}|=1} \tfrac{1}{2}(1+\bm{s}\cdot\bm{s}_\cE) 
\end{equation}
where $\bm{s}$ is the Bloch vector of the input state $\psi$, and $\bm{s}_\cE$
is that of the output $\cE(\psi)$.
For a unital qubit channel, $\bm{s}_\cE$ is the
image of a linear map on the Bloch vector: $\bm{s}_\cE=\bm{Ms}$.
The minimum fidelity can thus be written simply as
\begin{equation}\label{eq:Fmin}
  \Fmin = \min_{\bm{s}: |\bm{s}|=1}
  \tfrac{1}{2}{\left(1+\bm{s}^T\bm{Ms}\right)}=\frac{1}{2}(1+\mu_{\text{min}}),
\end{equation}
where $\mu_{\text{min}}$ is the smallest eigenvalue of
$\frac{1}{2}(\bm{M}+\bm{M}^\dagger)$.
This provides the direct connection between the unital qubit channel and
$\Fmin$, and, in particular, allows us to express $\Fmin$ in terms of the
tomographic probabilities associated with a channel $\cE$. 

\FigFour

Here, we assume the promise that the unknown channel is a unital one; the
Pauli channel used to simulated the data is indeed unital.
In effect, this unitality assumption restricts the relevant space of Choi
states dual to the channels, to a strict subset of $\STP$, namely, to those
that also satisfy Eq.~\eqref{eq:unital}.
Any channel sampling is thus done only from this subset.
Using the parameterization of Sec.~\ref{sec:UnitalParam}, we employ HMC
integration to compute the marginal likelihood $L(D|\Fmin)$.
The result is given in Fig.~\ref{fig:minMarginalLikelihood}; the intermediate
steps are provided in Appendix \ref{app:Supp2}.
With this marginal likelihood, one can construct the corresponding SCIs and
the plausible region, as well as perform other statistical inferences
about the unital qubit channel.

\subsection{Model selection}\label{sec:MS}
\TblOne
Often, one may not need the full generality of a CPTP channel to describe the
dynamics of a quantum system.
Instead, a simpler model with fewer parameters may suffice.
Simpler models are computationally easier to work with, are likely more easily
motivated from a physical standpoint, and may already describe the tomographic
data well.
One can phrase this problem as one of model selection in statistics, where the
best model, among a few candidate models, is chosen, given the available
data.
Here, we discuss the quantum problem of model selection for channel families.
Our sampling algorithm is used for two purposes here: (1) to evaluate a
criterion---based on the notion of relative belief---for the ``best'' model;
(2) to assess and compare the performance of different model selection
criteria by testing them on many randomly chosen true channels. 

Two criteria for model selection commonly used in classical problems are the
Akaike Information Criterion (AIC) \cite{AIC} and the Bayesian Information
Criterion (BIC) \cite{BIC}.
The AIC is based on the quantity (which we denote also as ``AIC''),
\begin{equation}
 \textrm{AIC} = 2k-2\log(L_{\text{max}}),
\end{equation}
where $k$ is the number of parameters in the model and $L_{\max}$ is the
maximum value of the likelihood of the model for the data.
The best model is the one with the smallest AIC value.
The BIC is defined in a similar manner, but uses the value of $N$, the number
of copies measured, 
\begin{equation}
\textrm{BIC} = k\log(N)-2\log(L_{\text{max}}).
\end{equation}
The best model according to this criterion is again the one with the smallest
BIC value. 

Another approach to model selection is based on the relative belief ratio
(RBR) of Ref.~\cite{Evans:15}.
The RBR of a model $M$ is the ratio of its posterior to prior probabilities,
\begin{equation}
  \RBR(M|D) = \frac{P(M|D)}{P(M)},
\end{equation}
where
\begin{equation}
P(M|D)=\!\int\limits_{M}\!\!(\upd\rho_\cE)
  \,\frac{L(D|\rho_\cE)}{L(D)}
\quad \textrm{and}\quad
P(M)=\!\int\limits_{M}\!\!(\upd\rho_\cE).
\end{equation}
If the posterior probability for a model $M$ increases after the data,
i.e. ${\RBR(M|D)>1}$, the data  provide evidence in favor of the model; the data
provide evidence against the model if ${\RBR(M|D)<1}$.    
It is also useful to have a measure of strength of evidence, since the data
might provide evidence in favor of more than one model from our candidate
set, and one would like some basis of choosing among those models.
The $\RBR$ value by itself is not a measure of the strength of evidence
(see Ref.~\cite{Evans:15} for a discussion of various aspects, and also
Ref.~\cite{Evans19}). 
We supplement it with the posterior probability
\begin{equation} 
P_{M_0}\equiv P\bigl([\RBR(M|D)=\RBR(M_0|D)]\,|\,D\bigr),
\end{equation}
for the model $M_0$ in question, and $M$ ranges over the set of candidate
models. 
If ${\RBR(M_0|D)>1}$ \emph{and} $P_{M_0}$ is large, then there is strong
evidence in favor of $M_0$.
The best model, according to the RBR criterion of relative belief ratio, is
the one with the largest posterior probability $P_M$, among all candidate
models with ${\RBR(M|D)>1}$.  

\begin{figure}
\centerline{\includegraphics[width=0.9\columnwidth]{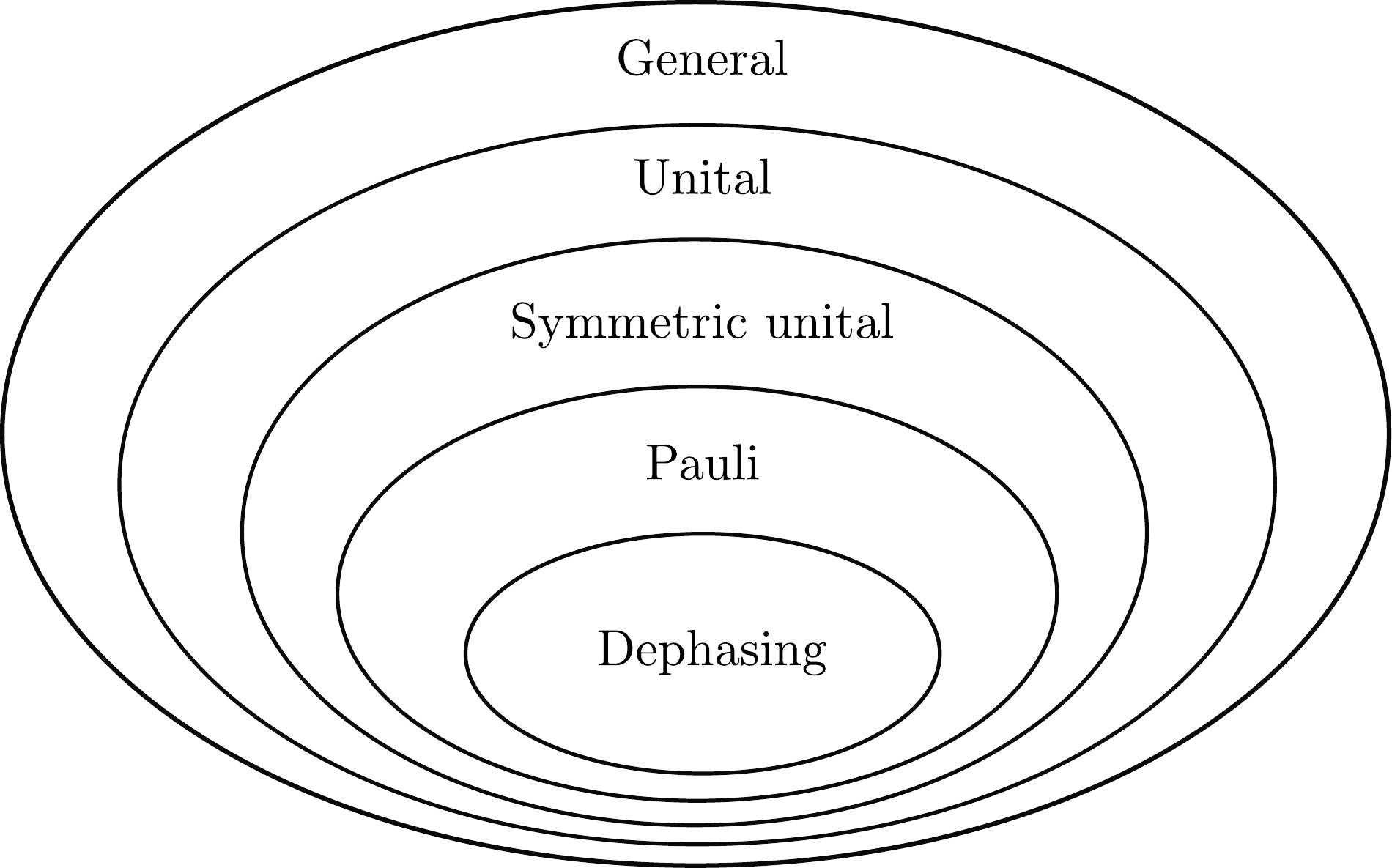}}
\caption{\label{fig:models}
The hierarchy of the five candidate models for the example in  Sec.~\ref{sec:MS}.} 
\end{figure}

As an example, we consider as candidate models five nested qubit channel
families: dephasing channels $\subset$ Pauli channels $\subset$ symmetric
unital channels $\subset$ unital channels $\subset$ general CPTP channels; see Fig.~\ref{fig:models}. The smallest set is the 1-parameter family of dephasing channels,
\begin{equation}
  \Bigl\{\cD_p(\cdot)\equiv (1-p)(\cdot)
            +p\sigma_z(\cdot)\sigma_z,p\in[0,1]\Bigr\},
\end{equation} 
and its Choi state is given by
\begin{equation}
\rho_\cE = {\left(\begin{array}{cccc}
1 & 0 & 0 & 1-2p \\
0 & 0 & 0 & 0 \\
0 & 0 & 0 & 0 \\
1-2p & 0 & 0 & 1
\end{array}\right)}.
\end{equation}
The set of Pauli channels is a 3-parameter family,
\begin{equation}
  {\left\{\textrm{Pauli}_{\mathbf{p}}(\cdot)
      \equiv {\Bigl(1-\sum_ip_i\Bigr)}(\cdot)
      +\sum_ip_i\sigma_i(\cdot)\sigma_i\right\}},
\end{equation}
for $\mathbf{p}\equiv (p_x,p_y,p_z)$, $p_i\geq 0$, and $\sum_ip_i\leq 1$.
The Choi state of a Pauli channel is of the form
\begin{equation}
\rho_\cE = {\left(\begin{array}{cccc}
p_I+p_z & 0 & 0 & p_I-p_z \\
0 & p_x+p_y & p_x-p_y & 0 \\
0 & p_x-p_y & p_x+p_y & 0 \\
p_I-p_z & 0 & 0 & p_I+p_z
\end{array}\right)},
\end{equation}
where $p_I=1-p_x-p_y-p_z$.
The 6-parameter family of symmetric unital channels refers to the subset of
unital qubit channels such that $\bm{R_1}=\bm{R_2}$ in
Eq.~\eqref{eq:dyadic}.
We then have the 9-parameter family of unital qubit channels, and lastly, the
12-parameter set of all CPTP qubit channels. 

\TblTwo

A natural prior on the model space is one that puts equal weights on each 
family.
This is easily defined by the sampling procedure: 
the prior sample is constructed by generating 500,000 sample points with the primitive prior for each family. For the dephasing channel, the sample is generated by sampling $p$ uniformly from $[0,1]$. For the Pauli channel, we obtain the sample by generating $(p_x,p_y,p_z)$ uniformly from the 3-simplex. For the symmetric unital, the unital, and the general channels, we make use of HMC and the parameterizations in Sec.~\ref{sec:param} to generate the sample points.
Note that in the numerical procedure that generates the samples for, say, the
set of Pauli channels, we will never come across a sample point that is
exactly a dephasing channel with ${p_x=0=p_y}$.
Thus, even though the channel families are nested sets, one can consider each
family to have prior probability of $\frac{1}{5}$.
We use this prior to compute the RBR criterion for simulated data of different
sizes. 
For our choice of prior, $P(M)=\frac{1}{5}$ for all models and $P(M|D)$ is calculated by taking the average of $\frac{L(D|\rho)}{L(D)}$ over the sample points for each model. $P_{M_0}$ is computed by summing the posterior probabilities $P(M|D)$ of all the models $M$ with the same RBR as model $M_0$. Typically, $P_{M_0}=P(M_0|D)$.

To assess the performance of the three model-selection criteria, for each
family of channels, we randomly (according to the primitive prior, as described above) draw 1000 channels.
For each channel, we simulate data---with tetrahedron input states and a
tetrahedron measurement [see Eqs.~\eqref{eq:TetraStates} and
\eqref{eq:TetraPOVM}]---for ${N=20}$, $50$, $100$, $1\,000$, 
$10\,000$, and $100\,000$ copies measured, and evaluate the AIC, BIC, and RBR
criteria for that data.   
Table \ref{tbl:ModelSelectionResult} shows the conclusions when the three
criteria are applied to the simulated data.  
When the number of measured copies is very small, i.e., $N=20$, the results
based on AIC and BIC show a strong bias towards simpler (i.e.,
fewer-parameters) models.
In particular, both criteria rarely identify the right model when the true
channel comes from the unital or general families.
Results based on RBR, however, show significantly more instances where the
correct model is identified for the more complex (i.e., more parameters)
models.
For a moderate number of measured copies, i.e. ${N=1\,000}$, AIC and RBR
give equally good results, whereas BIC shows a slight bias towards the simpler
models.
When the number of measured copies is very large, i.e., ${N=100\,000}$, results
based on BIC are most accurate whereas results based on AIC have a slight bias
to the more complex models. 
RBR also performs well in this regime.

Another aspect that we can check easily with our sampling procedure is the
bias in the prior.
This is particularly important for model selection based on the RBR criterion,
to be sure that the probability of drawing a wrong conclusion is low.
For example, for data that are typical for a unital channel, if we were to
conclude regularly that there is evidence in favor of the general CPTP
model and evidence against the unital model, there is bias in favor of
the general CPTP model and bias against the unital model.
To check for the bias, we draw 1000 random channels from each of the channel
families and simulate data based on these true channels.
The number of instances where the simulated data provide evidence against each
of the four candidate models are calculated.
The results are shown in Table~\ref{tbl:Bias}.
As can be seen from the table, there is no significant bias in the prior when
$N \geq 100$, and the bias decreases as the number of measured copies
increases.

\section{Conclusions}\label{sec:conc}
In this work, we constructed one exact parameterization for the space of CPTP
channels.
This parameterization has no superfluous parameters, and requires no
imposition of any added constraints.
These features make it possible to use the parameterization in a HMC
algorithm, for producing high-quality---in terms of low correlations---samples
of CPTP channels from a user-specified distribution.
We demonstrated the usefulness of our parameterization in sampling
applications taken from quantum process tomography.
The method applies to general quantum channel sampling problems. 

While our parameterization serves the purpose, it is, of course, just one of the many parameterizations that could be used in a HMC algorithm for sampling from the quantum channel space. For example, it is conceivable that a useful parameterization of a channel can be given in terms of the marginals and the copula of the respective Choi state \cite{copula}. This is unexplored territory.

A useful extension of this work will be to discover also an exact
parameterization for the case of CPTP and unital channels.
As discussed above, this additional requirement of unitality presents
difficulties that can be easily overcome only in the qubit situation.
The parameterization for the space of CPTP, unital channels beyond the qubit
case, remains an open problem.
Note that such a parameterization will give also a possibly useful description
of the space of all bipartite mixed quantum states with completely mixed
states on both the single-party states; our current parameterization gives the
larger space of states where only one of the two single-party states is
completely mixed.

\acknowledgments
This work is supported in part by the Ministry of Education, Singapore
(through grant number MOE2016-T2-1-130).
HKN is also supported by Yale-NUS College (through a start-up grant).
The Centre for Quantum Technologies is a Research Centre of Excellence funded
by the Ministry of Education and the National Research Foundation of
Singapore.

\appendix

\section{Hamiltonian Monte Carlo (HMC)}
\label{app:HMC}
HMC makes use of pseudo-Hamiltonian dynamics in a mock phase space. The parameters of interest are identified as the position variables $\theta$ and fictitious momentum variables $\vartheta$ are introduced. The Hamiltonian is defined as
\begin{equation}
H(\theta,\vartheta)=\frac{1}{2}\sum_j\vartheta_j^2 -\log w(\theta),
\end{equation}
where $w(\theta)$ is the target distribution. Any reasonable target distribution is permitted and, therefore, one can sample in accordance with any $w(\theta)$.

The HMC algorithm generates a set of sample points which follows the target distribution $w(\theta)$. The HMC algorithm is stated as follows \cite{Seah+1:15}:
\begin{center}
\textbf{HMC algorithm} 
\end{center}
\begin{enumerate}
\item Set $j=1$ and choose an arbitrary starting point $\theta^{(1)}$.
\item Generate $\vartheta^{(j)}$ from a multivariate Gaussian distribution with mean zero and unit variance.
\item Solve the Hamiltonian equations of motion
\begin{equation} \label{eq: Hamilton eq}
\frac{\mathrm{d}}{\mathrm{d}t}\theta_i=\frac{\partial}{\partial \vartheta_i} H, \quad \frac{\mathrm{d}}{\mathrm{d}t}\vartheta_i=-\frac{\partial}{\partial \theta_i} H
\end{equation}
with the initial conditions $(\theta,\vartheta)|_{t=0}=(\theta^{(j)},\vartheta^{(j)})$ to obtain $(\theta^*,\vartheta^*)=(\theta,-\vartheta)|_{t=T}$.
\item Calculate the acceptance ratio
\begin{equation}
a=\min\{1,e^{H(\theta^{(j)},\vartheta^{(j)})-H(\theta^*,\vartheta^*)}\}.
\end{equation}
\item Draw a random number $b$ uniformly from $[0,1]$. If $b<a$, set $\theta^{(j+1)}=\theta^*$; otherwise, set $\theta^{(j+1)}=\theta^{(j)}$.
\item Set $j=j+1$. If $j$ equals the desired number of samples, escape the loop; otherwise, return to step 2.
\end{enumerate}

Note that the $\vartheta$ distribution in step 2 is proportional to the kinetic-energy factor in $\upe^{-H(\theta,\vartheta)}$; the HMC algorithm enforces a $\theta$ distribution proportional to the potential-energy factor in $\upe^{-H(\theta,\vartheta)}$, which is $\upe^{\log w(\theta)}=w(\theta)$, the target distribution. If the differential equations in \eqref{eq: Hamilton eq} can be solved exactly, then the acceptance ratio $a=1$. In practice, the differential equations must be discretized. This is done by the leapfrog method. Due to the discretization error, the acceptance ratio will not be 1 generally. The leapfrog method should be implemented such that the acceptance ratio is around the optimal value of $65\%$ \cite{Neal11}.

\section{Parameterizing qutrit channels}\label{app:Qutrit}
Here, we report an explicit application of the
parameterization of Sec.~\ref{sec:param}, for the case of qutrit channels.
We start with the permutation matrix $P$ that reshuffles $\varphi_i$s into
$\psi_i$s, with the identically zero entries located below the generically
nonzero ones. A $P$ that can accomplish this is one such that\par 
\begin{equation}
P
\begin{pmatrix}
\begin{smallmatrix}\\
  1\\2\\3\\4\\5\\6\\7\\8\\9\\10\\11\\12\\13\\14\\15\\16\\17\\18\\19\\
  20\\21\\22\\23\\24\\25\\26\\27\\
\end{smallmatrix}
\end{pmatrix}
=
\begin{pmatrix}
\begin{smallmatrix}\\
  1\\10\\11\\19\\20\\21\\2\\3\\4\\12\\13\\14\\22\\23\\24\\5\\6\\
  7\\15\\16\\17\\25\\26\\27\\8\\9\\18\\
\end{smallmatrix}
\end{pmatrix}.
\end{equation}
After the permutation, we have
\begin{equation}
\psi_1 = \begin{pmatrix}
\psi_{1,1} \\
\vdots \\
\psi_{1,6} \\
0 \\
\vdots \\
0
\end{pmatrix},
\quad
\psi_2 = \begin{pmatrix}
\psi_{2,1} \\
\vdots \\
\psi_{2,15} \\
0 \\
\vdots \\
0
\end{pmatrix},
\quad
\psi_3 = \begin{pmatrix}
\psi_{3,1} \\
\vdots \\
\psi_{3,24} \\
0 \\
0 \\
0
\end{pmatrix}.
\end{equation}
To parameterize the $\psi_i$s such that they are orthonormal, we first
parameterize $\psi_3$, of unit length, 
\begin{equation} 
\psi_3 = \begin{pmatrix}
\Expi{\phi_1}\sin\theta_1 \sin\theta_2 \cdots \sin\theta_{22} \sin\theta_{23}  \\
\Expi{\phi_2}\cos\theta_1 \sin\theta_2 \cdots \sin\theta_{22} \sin\theta_{23}  \\
\vdots \\
\Expi{\phi_{23}}\cos\theta_{22} \sin\theta_{23} \\
\Expi{\phi_{24}}\cos\theta_{23}  \\
0 \\
0 \\
0
\end{pmatrix}.
\end{equation}
Recalling that $\varphi_{33}$ [see Eq.~\eqref{eq:AMatrix}] is a $d^2$-entry
real column, and with the $P$ given above, $\psi_{3,4}$, $\psi_{3,5}$,
$\psi_{3,6}$, $\psi_{3,13}$, $\psi_{3,14}$, $\psi_{3,15}$, $\psi_{3,22}$,
$\psi_{3,23}$, $\psi_{3,24}$ are real. 
Thus, $\phi_4$, $\phi_5$, $\phi_6$, $\phi_{13}$, $\phi_{14}$, $\phi_{15}$,
$\phi_{22}$, $\phi_{23}$, $\phi_{24}$ are set to zero.
Then, we define $\{v_n\}_{n=1}^{14}$ which lie in the orthogonal subspace of
$\psi_3$ as follows, 
\begin{align}
v_1 &= \begin{pmatrix}
\Expi{\phi_1}\cos\theta_1 \\
-\Expi{\phi_2}\sin\theta_1 \\
0 \\
\vdots \\
0
\end{pmatrix},
\quad
v_2 = \begin{pmatrix}
\Expi{\phi_1}\sin\theta_1 \cos\theta_2 \\
\Expi{\phi_2}\cos\theta_1 \cos\theta_2 \\
-\Expi{\phi_3}\sin\theta_2 \\
0 \\
\vdots \\
0
\end{pmatrix}, 
~\cdots,\nonumber\\
v_{14} &= \begin{pmatrix}
\Expi{\phi_1}\sin\theta_1 \sin\theta_2 \cdots \sin\theta_{13} \cos\theta_{14}  \\
\Expi{\phi_2}\cos\theta_1 \sin\theta_2 \cdots \sin\theta_{13} \cos\theta_{14}  \\
\vdots \\
\Expi{\phi_{14}}\cos\theta_{13} \cos\theta_{14} \\
-\Expi{\phi_{15}}\sin\theta_{14}  \\
0 \\
\vdots \\
0
\end{pmatrix}.
\end{align}
To make $\psi_1$ and $\psi_2$ orthogonal to $\psi_3$, we set them to be in the
span of $\{v_n\}$, 
\begin{align}
\begin{pmatrix}
| &  | \\
\psi_1  & \psi_2  \\
| & |
\end{pmatrix}&\equiv V\widetilde\Psi 
= \begin{pmatrix}
| & | &  & | \\
v_1     &  v_2 & ...  & v_{14} \\
| & | & & |
\end{pmatrix}
\begin{pmatrix}
| &  | \\
\widetilde\psi_1  & \widetilde\psi_2  \\
| & |
\end{pmatrix}
\end{align}
The orthonormality of $\psi_1$ and $\psi_2$ is equivalent to the
orthonormality of $\widetilde\psi_1$ and $\widetilde\psi_2$.
We simply need to repeat the previous procedure.
We parameterize $\widetilde\psi_2$ to be of unit length,
\begin{equation}
\widetilde\psi_2 = \begin{pmatrix}
\Expi{\widetilde{\phi}_1}\sin\widetilde{\theta}_1 \sin\widetilde{\theta}_2
\cdots \sin\widetilde{\theta}_{12} \sin\widetilde{\theta}_{13}  \\ 
\Expi{\widetilde{\phi}_2}\cos\widetilde{\theta}_1 \sin\widetilde{\theta}_2
\cdots \sin\widetilde{\theta}_{12} \sin\widetilde{\theta}_{13}  \\ 
\vdots \\
\Expi{\widetilde{\phi}_{13}}\cos\widetilde{\theta}_{12}
\sin\widetilde{\theta}_{13} \\ 
\Expi{\widetilde{\phi}_{14}}\cos\widetilde{\theta}_{13} \\
\end{pmatrix}.
\end{equation}
Next, we define $\{u_n\}_{n=1}^{4}$, each orthogonal to $\psi_2$,
\begin{align}
u_1 &= \begin{pmatrix}
\Expi{\widetilde{\phi}_1}\cos\widetilde{\theta}_1 \\
-\Expi{\widetilde{\phi}_2}\sin\widetilde{\theta}_1 \\
0 \\
\vdots \\
0
\end{pmatrix},
\qquad
u_2 = \begin{pmatrix}
\Expi{\widetilde{\phi}_1}\sin\widetilde{\theta}_1 \cos\widetilde{\theta}_2 \\
\Expi{\widetilde{\phi}_2}\cos\widetilde{\theta}_1 \cos\widetilde{\theta}_2 \\
-\Expi{\widetilde{\phi}_3}\sin\widetilde{\theta}_2 \\
0 \\
\vdots \\
0
\end{pmatrix}, 
~\ldots,\nonumber\\
u_{4} &= \begin{pmatrix}
\Expi{\widetilde{\phi}_1}\sin\widetilde{\theta}_1 \sin\widetilde{\theta}_2
\sin\widetilde{\theta}_{3} \cos\widetilde{\theta}_{4}  \\ 
\Expi{\widetilde{\phi}_2}\cos\widetilde{\theta}_1 \sin\widetilde{\theta}_2
\sin\widetilde{\theta}_{3} \cos\widetilde{\theta}_{4}  \\ 
\Expi{\widetilde{\phi}_{4}}\cos\widetilde{\theta}_{2}
\sin\widetilde{\theta}_{3} \cos\widetilde{\theta}_{4} \\ 
\Expi{\widetilde{\phi}_{4}}\cos\widetilde{\theta}_{3}\cos\widetilde{\theta}_{4}
\\ 
-\Expi{\widetilde{\phi}_{5}}\sin\widetilde{\theta}_{4} \\
0 \\
\vdots \\
0
\end{pmatrix}.
\end{align}
Finally, to have $\widetilde{\psi}_1$ normalized and orthogonal to
$\widetilde{\psi}_2$, we set 
\begin{align}
\widetilde{\psi}_1&\equiv U\bar{\psi}_1 = \begin{pmatrix}
| & | &  & | \\
u_1     &  u_2 & ...  & u_4 \\
| & | & & |
\end{pmatrix} \bar{\psi}_1,
\end{align}
where 
\begin{equation}
\bar{\psi}_1 = \begin{pmatrix}
\Expi{\bar{\phi}_1}\sin\bar{\theta}_1 \sin\bar{\theta}_2 \sin\bar{\theta}_{3}
\\ 
\Expi{\bar{\phi}_2}\cos\bar{\theta}_1 \sin\bar{\theta}_2  \sin\bar{\theta}_{3}
\\ 
\Expi{\bar{\phi}_{3}}\cos\bar{\theta}_{2} \sin\bar{\theta}_{3} \\
\Expi{\bar{\phi}_{4}}\cos\bar{\theta}_{3}
\end{pmatrix}.
\end{equation}

We check that we have the right number of parameters.
The parameters used above are $\theta_1,...,\theta_{23}$,
$\phi_1,...,\phi_{24}$ (nine of these are set identically to zero),
$\widetilde{\theta}_1,...,\widetilde{\theta}_{13}$,
$\widetilde{\phi}_1,...,\widetilde{\phi}_{14}$,
$\bar{\theta}_1,\bar{\theta}_2,\bar{\theta}_3$, and
$\bar{\phi}_1,...,\bar{\phi}_4$, giving a total of $72=3^2(3^2-1)$ parameters,
as needed for specifying qutrit channels.

\section{Iterative algorithm for  estimating the  marginal likelihood}
\label{app:Iter} 
To estimate the marginal likelihood reliably, we follow the procedure in
Ref.~\cite{OEI16}.
For the following discussion, we assume  
\begin{equation}
  0\leq f(p) \leq 1,
\end{equation} 
for the sake of simplicity.
First, we note that the integrands in \eqref{eq:marlik1} are ill-suited for
MC integration due to the presence of the Dirac delta factors.
We consider the antiderivatives
\begin{equation}
  P_{r,0}(F) = \int (\mathrm{d}p)\,w_r(p)\eta\bigl(F-f(p)\bigr)
\end{equation}
and
\begin{equation}
  P_{r,D}(F) = \frac{1}{L(D)}\int (\mathrm{d}p)\,w_r(p)
  \eta\bigl(F-f(p)\bigr)L(D|p).
\end{equation} 
With a sample of $w_r(p)$ and $\sfrac{w_r(p)L(D|p)}{L(D)}$, we can evaluate the
antiderivatives for various values of $F$ and fit them with several-parameters
functions.
From the fitted functions, we can then calculate the derivatives
\begin{equation}
  W_{r,0}(F) = \frac{\partial}{\partial F}P_{r,0}(F)
            = \int (\mathrm{d}p)\,w_r(p)\delta\bigl(F-f(p)\bigr)
\end{equation}
and
\begin{align}
  W_{r,D}(F) &= \frac{\partial}{\partial F}P_{r,D}(F)  \nonumber\\
             &= \frac{1}{L(D)}\int(\mathrm{d}p)\,w_r(p)
                 \delta\bigl(F-f(p)\bigr)L(D|p)
\end{align}
and obtain the marginal likelihood by
\begin{equation} \label{eq:marlik2}
  L(D|F) = \frac{W_{r,D}(F)}{W_{r,0}(F)}.
\end{equation}

A problem arises when $P_{r,0}(F)$ is very close to a constant over some range
of values of $F$. 
The common situation is that $P_{r,0}(F)$ is very close to zero for a range of
values near ${F=0}$ and very close to one for a range of values near ${F=1}$.
MC integration is not precise enough to distinguish ${P_{r,0}(F)\gtrsim0}$ from
${P_{r,0}(F)=0}$ and ${P_{r,0}(F)\lesssim1}$ from ${P_{r,0}(F)=1}$.
As a result, the estimated value of $W_{r,0}(F)$ will be equal to zero over
those range of values.
We cannot get a reliable estimation of $L(D|F)$ in this situation since
$W_{r,0}(F)$ is the denominator in Eq.~\eqref{eq:marlik2}.
To overcome this problem, we note that we can do the replacement
\begin{equation}
   w_r(p) \rightarrow w_r(p)g\bigl(f(p)\bigr)
\end{equation} 
with an arbitrary function ${g(F)>0}$ without changing the value of $L(D|F)$. 

The procedure for obtaining a reliable estimation of $L(D|F)$ is as follows:
\begin{enumerate}
\item Sample according to $w_r(p)$.
  Use this sample to calculate $P_{r,0}(F)$.
  Fit a several-parameters function to $P_{r,0}(F)$ and obtain $W_{r,0}(F)$ by
  differentiating the fitted function. 
\item Sample according to
  $\widetilde{w}_r(p)=\sfrac{w_r(p)}{W_{r,0}\bigl(f(p)\bigr)}$. 
  Use this sample to calculate
\begin{equation}
  \widetilde P_{r,0}(F) = \int (\mathrm{d}p)\,\widetilde w_r(p)\eta\bigl(F-f(p)\bigr).
\end{equation}
Fit a several-parameters function to $\widetilde P_{r,0}(F)$ and obtain
$\widetilde W_{r,0}(F)$ by differentiating the fitted function. 
\item Sample according to $\sfrac{\widetilde
    w_r(p)L(D|p)}{L(D)}=\sfrac{w_r(p)L(D|p)}{W_{r,0}\bigl(f(p)\bigr)L(D)}$.
  Use this sample to calculate
\begin{equation}\qquad\quad
  \widetilde P_{r,D}(F) = \int (\mathrm{d}p)\widetilde w_r(p)
  \eta\bigl(F-f(p)\bigr)L(D|p).
\end{equation}
Fit a several-parameters function to $\widetilde P_{r,D}(F)$ and obtain $\widetilde
W_{r,D}(F)$ by differentiating the fitted function. 
\item Obtain the marginal likelihood from
\begin{equation}
L(D|F) = \frac{\widetilde W_{r,D}(F)}{\widetilde W_{r,0}(F)}.
\end{equation}
\end{enumerate}

The reason that we can have a reliable estimation of $L(D|F)$ using $\widetilde
W_{r,0}(F)$ obtained in step~2 is as follows.
Suppose the exact value of $W_{r,0}(F)$ is known, $\widetilde W_{r,0}(F)$ will be
equal to $1$ and $\widetilde P_{r,0}(F)$ will be equal to $F$.
If the exact values of $W_{r,0}(F)$ are not known, but we have
a good approximation for $W_{r,0}(F)$ from step~1 and use it for the
calculation of $\widetilde P_{r,0}(F)$ in step~2, the $\widetilde P_{r,0}(F)$ that we
obtain will still be quite close to $F$ and $\widetilde W_{r,0}(F)$ will be
nonzero for all range of $F$ values. 

\FigSix\FigSeven

In step~1, $P_{r,0}(F)$ can be fitted with a linear combination of regularized
incomplete beta functions 
\begin{equation}
  I_{a,b}(x) = \frac{\int_0^x t^{a-1}(1-t)^{b-1}\mathrm{d}t}
                   {\int_0^1 t^{a-1}(1-t)^{b-1}\mathrm{d}t},
\end{equation}
that is
\begin{align} \label{eq:RIBFfit}
P_{r,0}(F) = &w_1I_{a_\text{min},b_1}(F) + w_2I_{a_1,b_\text{min}}(F) \nonumber\\
&+ w_3I_{a_2,b_2}(F) +\dots  \nonumber\\
&+ \left(1-\sum_{i=1}^{N-1}w_i\right)I_{a_{N-1},b_{N-1}}(F),
\end{align}
with the fitting parameters $a_1,\dots,a_{N-1},$ $b_1,\dots,b_{N-1},$
$w_1,\dots,w_{N-1}$.
$a_\text{min}$ and $b_\text{min}$ are fixed by the power laws satisfy by
$P_{r,0}(F)$ near $F=0$ and $F=1$,  
\begin{equation}
P_{r,0}(F) \propto F^{a_\text{min}} \quad \text{for} \; F\gtrsim0,
\end{equation}
and
\begin{equation}
1 - P_{r,0}(F) \propto (1-F)^{b_\text{min}} \quad \text{for} \; F\lesssim1.
\end{equation}
In step~2, a truncated Fourier series of the form 
\begin{align}
\widetilde P_{r,0}(F) \simeq F &+ c_1\sin(\pi F)+ c_2\sin(2\pi F) \nonumber\\
&+ c_3\sin(3\pi F) + \cdots
\end{align}
is usually a good fitting function.
In step~3, $\widetilde P_{r,D}(F)$ can be fitted with a smoothing spline.

\section{Intermediate results for the estimation of  the marginal likelihood}
\label{app:Supp}
\subsection{Average gate fidelity}\label{app:Supp1}
The green dots in Fig.~\ref{fig:avgGateFid}(a) show the values of
$P_{r,0}(\Favg)$ obtained by a MC integration with $1\,000\,000$ sample
points. 
The MC integration is not precise enough to distinguish
${P_{r,0}(\Favg)\gtrsim0}$ from ${P_{r,0}(\Favg)=0}$ near
${\Favg=\frac{1}{3}}$ and to distinguish ${P_{r,0}(\Favg)\lesssim1}$
from ${P_{r,0}(\Favg)=1}$ near ${\Favg=1}$.
Therefore, a reliable approximation for $W_{r,0}(\Favg) =
\sfrac{\partial}{\partial \Favg}P_{r,0}(\Favg)$ cannot be
obtained.
To overcome this problem, we follow the procedure stated in
Appendix~\ref{app:Iter}. 
First, we fit the green dots with a three-term fitting
function of the form of 
Eq.~\eqref{eq:RIBFfit} with ${F=\frac{3}{2}(\Favg-\frac{1}{3})}$,
${a_{\text{min}}=3}$ and ${b_{\text{min}}=\frac{21}{2}}$.
The black curve is the fitted curve of $P_{r,0}(\Favg)$.
The fitting parameters are shown in the inset table. 
$\widetilde P_{r,0}(\Favg)$ is obtained from a MC integration with
$1\,500\,000$ sample points and shown as the blue dots in
Fig.~\ref{fig:avgGateFid}(a). 

The $\widetilde P_{r,0}(\Favg)$ is quite close to the straight line
${\frac{3}{2}(\Favg-\frac{1}{3})}$.
The $\widetilde P_{r,0}(\Favg)$ after subtracting the straight line
${\frac{3}{2}(\Favg-\frac{1}{3})}$ is shown as the blue dots in
Fig.~\ref{fig:avgGateFid}(b). 
The blue curve shows the fitting curve, a truncated Fourier series whose
Fourier amplitudes are reported in Fig.~\ref{fig:avgGateFid}(c).

$\widetilde P_{r,D}(\Favg)$ is evaluated by a MC integration with $1\,500\,000$
sample points and it can be fitted with a smoothing spline.
The marginal likelihood shown in
Fig.~\ref{fig:avgMarginalLikelihood} is obtained from the
ratio of $\widetilde W_{r,D}(\Favg)$ and $\widetilde W_{r,0}(\Favg)$.

\subsection{Worst-case fidelity  of a unital qubit channel}\label{app:Supp2}
The green dots in Fig.~\ref{fig:minGateFid}(a) show the values of
$P_{r,0}(\Fmin)$ from a MC integration with $1\,000\,000$ sample points.
The MC integration is not precise enough to distinguish
${P_{r,0}(\Fmin)\gtrsim0}$ from ${P_{r,0}(\Fmin)=0}$ near ${\Fmin=0}$ and
to distinguish ${P_{r,0}(\Fmin)\lesssim1}$ from ${P_{r,0}(\Fmin)=1}$ near ${\Fmin=1}$.
Therefore, a reliable approximation for ${W_{r,0}(\Fmin) =
\sfrac{\partial}{\partial \Fmin}P_{r,0}(\Fmin)}$ cannot be
obtained.
To overcome this problem, we follow the procedure stated in
Appendix~\ref{app:Iter}.
First, we fit the green dots with a three-term fitting function of the form 
in Eq.~\eqref{eq:RIBFfit} with ${F=\Fmin}$,
${a_{\text{min}}=4}$, and ${b_{\text{min}}=\frac{15}{2}}$.
The black curve is fitted to the numerical values for $P_{r,0}(\Fmin)$.
The fitting parameters are shown in the inset table.
$\widetilde P_{r,0}(\Fmin)$ is obtained by a MC integration with $1\,500\,000$
sample points and shown as the blue dots in Fig.~\ref{fig:minGateFid}(a).

The values of $\widetilde P_{r,0}(\Fmin)$ are quite
close to the straight line $\Fmin$.
The corresponding values after
subtracting this straight line make up the blue dots in
Fig.~\ref{fig:minGateFid}(b). 
The blue fitting curve is a truncated Fourier series with the Fourier
amplitudes of Fig.~\ref{fig:minGateFid}(c). 

$\widetilde P_{r,D}(\Fmin)$ is evaluated by a MC integration with $1\,500\,000$
sample points and it can be fitted with a smoothing spline.
The marginal likelihood shown in Fig.~\ref{fig:minMarginalLikelihood} is the
ratio of $\widetilde W_{r,D}(\Fmin)$ and $\widetilde W_{r,0}(\Fmin)$.


\begin{thebibliography}{99}
\bibitem{MagicBasis}
An alternative, equally straightforward, parameterization exploits that the
two-qubit Choi states for unital channels have real $4\times4$ matrices in
the ``magic basis'' \cite{Hill-Wootters97}, composed of the singlet state
$\ket{\mathrm{s}}=2^{-\frac{1}{2}}\bigl(\ket{01}-\ket{10}\bigr)$ and the three
triplet states $\mathrm{i}\sigma_x\otimes\id\ket{s}$,
$\mathrm{i}\sigma_y\otimes\id\ket{s}$,
and $\mathrm{i}\sigma_z\otimes\id\ket{s}$.
\end{thebibliography}
\end{document}